\documentclass[10pt,journal,compsoc]{IEEEtran}

\usepackage{cite}
\usepackage{amsmath,amssymb,amsfonts}
\usepackage{graphicx}
\usepackage{textcomp}

\usepackage{amssymb} 
\usepackage{comment}
\usepackage{xcolor}
\usepackage[colorlinks=true]{hyperref}
\usepackage{subcaption}

\usepackage{amsthm}
\usepackage[T1]{fontenc}
\usepackage{newtxtext,newtxmath}
\usepackage{algorithm}
\usepackage{algpseudocode}
\usepackage{forest}
\usepackage{bbding}

\usepackage{multicol,multirow}

\usepackage{array}
\usepackage{makecell}

\newcolumntype{C}[1]{>{\centering\arraybackslash}m{#1}}

\newtheorem{proposition}{Proposition}
\newtheorem{corollary}{Corollary}
\newtheorem{lemma}{Lemma}

\newtheorem{assumption}{Assumption}

\begin{document}

\title{Reactive Users vs. Social Recommender Systems: Managing Opinion Drifts with\\ Adaptive Policies}
\author{Atefeh Mollabagher and Parinaz Naghizadeh%
\thanks{A. Mollabagher and P. Naghizadeh are with the Department of Electrical and Computer Engineering, University of California, San Diego, CA 92093 USA. E-mail: \texttt{\{atefeh,\,parinaz\}@ucsd.edu}.}%
\thanks{This work is supported in part by the NSF under award \#2411000.}%
}

\maketitle
\begin{abstract}
Recommendation systems are used in a range of platforms to maximize user engagement through personalization, promotion of popular content, and the use of information from social networks. It has been found that such recommendations may shape users' opinions over time. In this paper, we ask whether \emph{reactive} users, who are cognizant of the influence of the content they consume, can limit such changes by adaptively adjusting their content consumption choices. To this end, we study users' opinion dynamics under two stochastic content consumption policies: a \emph{passive} policy, where the probability of clicking on recommended content is fixed, and a \emph{reactive} policy, where the probability of content consumption adaptively decreases following large opinion drifts. We analytically derive the expected opinion and user utility under these policies when a user is influenced by both a social network and the recommender. We show that the adaptive policy can help users prevent opinion drifts induced by recommendations and that when a user prioritizes opinion preservation, the expected utility of the adaptive policy can outperform the fixed policy. We validate our theoretical findings through numerical simulations. These findings help better understand how user-level strategies can challenge the biases induced by recommendation systems.
\end{abstract}

\begin{IEEEkeywords}
Opinion Dynamics, Recommendation Systems, Reactive Users, Social Network.
\end{IEEEkeywords}

 
\section{Introduction}\label{sec:introduction}

Recommendation systems are used across a variety of online platforms, including streaming services, online shopping, social networks, and news aggregators. They efficiently process vast amounts of information and uncover patterns that help users find what they are more likely to enjoy, more easily. Social recommender systems, in particular, use the underlying social networks to better learn users' preferences, and further, to make recommendations more persuasive and reliable by leveraging users' social influence and trust. It has been shown that users tend to closely follow the recommendations they receive, even at times when these recommendations are random~\cite{ursu2018power}.
This (over-)reliance on recommendation systems carries negative effects, one of which is undesired changes in users' preferences and behavior over time. For instance: (i) repeated exposure to certain content through social networks can lead users to adopt extreme opinions~\cite{allcott2020welfare,levy2021social}, (ii) exposure to personalized recommendations can amplify users' preexisting beliefs~\cite{terren2021echo}, a phenomenon known as ``echo chambers'' in the context of news/social media, and (iii) repeated exposure to similar opinions and viewpoints can isolate users from opposing views, creating so-called ``filter bubbles''~\cite{pariser2011filter}.

The current prevalent approach to address such undesired influence of (social) recommender systems on users is through digital platform regulations and content moderation systems (e.g.,  \cite{di2022recommender}). In this paper, we explore a complementary perspective and ask whether users can use their agency, \emph{themselves}, and adaptively adjust their engagement with recommended content. We refer to such agents as \emph{reactive users}, who sometimes intentionally ignore recommendations to prevent substantial changes in their innate preferences. This is in contrast to prior work on \emph{passive users} who consume, and are influenced by, all recommended content \cite{rossi2021closed, lanzetti2023impact}. 

Our motivation to explore reactive users' ability to manage opinion drifts is two-fold: (i) recent surveys \cite{deanaccounting} show that users indeed adopt strategic content consumption decisions when interacting with recommendations, and (ii) psychologically grounded models of agents' opinion dynamics \cite{curmei2022towards}, and specifically, ``hedonic adaptation'', argue that users return to a baseline opinion after some time, even if there are temporary changes in one's opinion; as such, decreasing the consumption of recommended content should allow users to return to their innate preferences.

Formally, we begin by proposing a model of opinion dynamics arising from the interactions between a recommendation system (platform) and \emph{reactive} users in a social network. Our model extends those of \cite{rossi2021closed,lanzetti2023impact, sprenger2024control}, which analyze the opinion dynamics of \emph{passive} users who consume, and are always influenced by, the recommended content. In contrast, we introduce a model in which users follow one of two stochastic content consumption policies: a baseline \emph{fixed} content consumption policy (which includes the passive agents' policy of prior works as a special case), and an \emph{adaptive decreasing} policy (which captures reactive users who adjust their use of the platform adaptively so as to control their opinion drift). For each policy, we characterize the short-term and long-term expected opinions (allowing us to assess the extent of opinion drift) and expected agent utilities (allowing us to compare the agent's preference for each policy). 

Our first main finding is that, in a mixed population where all other agents remain passive, the adaptive decreasing policy enables a reactive agent to better limit its expected opinion drift than a fixed policy. This is because the adaptive decreasing policy reduces content consumption when needed. Formally, we show (Proposition~\ref{prop:gamma-convergence}) that, as long as recommendation and neighbors' opinions are sufficiently far from the reactive agent's innate opinion, content consumption probability under the adaptive policy converges to zero almost surely (we establish this using the Borel-Cantelli Lemma). As a result, in the long run, the recommendation no longer directly affects the reactive agent's opinion. This means that while the platform, who derives utility from content consumption, prefers passive behavior, users can leverage their agency and reduce content consumption to manage their opinion drifts.

Our second main finding is sufficient conditions (on the relative weight placed by the reactive agent on content consumption vs. opinion drifts) under which a reactive agent prefers the adaptive policy over the fixed policy, in both the short-term and the long-term, despite the reduced content consumption. The main challenge in this analysis comes from the path dependence of the adaptive policy and the threshold-triggered nature of the clicking probability reductions. This makes exact utility evaluation difficult, since both the expected drift and the expected click rate depend on the \emph{timing} of each click rate reduction, which in turn depends on the realizations of the click trajectories. We address this challenge (in Proposition~\ref{prop:finite-adaptive-better-than-fixed}) by bounding the two parts of the utility difference between the fixed and adaptive policies: the expected loss in content consumption under the adaptive policy, and the expected reduction in opinion drift.

To summarize, our main contributions in this paper are as follows: (i) we propose a new opinion dynamics model which accounts for the possibility of users deliberately rejecting some recommended content to limit their opinion drift, (ii) we propose an adaptive content consumption policy to model \emph{reactive} agents (in contrast to \emph{passive} agents considered in prior works), and (iii) we, analytically and numerically, show that a reactive policy enables users to manage opinion drift, and identify conditions under which reactive users following this policy can outperform passive users who adopt fixed content consumption decisions in a similarly passive population.
\section{Related Work}\label{sec:related-work}

\begin{table*}[!htbp]
\centering
\footnotesize
\setlength{\tabcolsep}{2.5pt}

\begin{tabular}{|C{1.55cm}||C{4.2cm}|C{3.4cm}|C{1.87cm}|C{3.1cm}|C{2.2cm}|}
\hline
Category 
& Related works
& \makecell{Opinion Dynamics?}
& \makecell{Social Network?}
& \makecell{Agent Click/Rating\\ Decisions?}
& \makecell{Strategic/Adaptive\\ Agents?}\\
\hline
\multirow{3}{=}{\centering Human-Rec Sys Feedback Loops}
& \makecell{Sinha et al. \cite{sinha2016deconvolving}, Krauth et al. \cite{krauth2025breaking}
}
& \XSolidBold & \XSolidBold & \CheckmarkBold & \XSolidBold \\
& Schmit et al. \cite{schmit2018human}
& \XSolidBold & \XSolidBold & \CheckmarkBold (deterministic selection) & \XSolidBold\\
& Haupt et al. \cite{haupt2023recommending}, Cen et al. \cite{cen2024user}, Cen et al. \cite{cen2024measuring}
& \XSolidBold & \XSolidBold & \CheckmarkBold & \CheckmarkBold(strategic) \\
\hline\hline

\multirow{2}{=}{\centering Opinion Drifts in Rec Sys}
& Curmei et al. \cite{curmei2022towards}, Lanzetti et al. \cite{lanzetti2023impact}
& \CheckmarkBold & \XSolidBold & \XSolidBold & \XSolidBold\\
& \makecell{
Rossi et al. \cite{rossi2021closed}, Jiang et al. \cite{jiang2019degenerate},\\
Kalimeris et al. \cite{kalimeris2021preference}
}
& \makecell{\CheckmarkBold(click-independent)}
& \XSolidBold & \CheckmarkBold & \XSolidBold\\
\hline\hline

\multirow{2}{=}{\centering Social Rec Sys}
& Castro et al. \cite{castro2017opinion}
& \makecell{\CheckmarkBold(DeGroot-like)}
& \CheckmarkBold & \XSolidBold & \XSolidBold \\
& \makecell{
Chandrasekaran et al. \cite{chandrasekaran2025network},\\
Sprenger et al. \cite{sprenger2024control}
}
& \makecell{\CheckmarkBold(FJ-like, click-independent)}
& \CheckmarkBold & \CheckmarkBold & \XSolidBold \\
\hline\hline

\multicolumn{2}{|c|}{This work}
& \makecell{\CheckmarkBold(FJ-like, click-dependent)}
& \CheckmarkBold
& \CheckmarkBold
& \makecell{\CheckmarkBold(adaptive)} \\
\hline
\end{tabular}
\caption{Summary of comparison with related work.}\label{table:comparison}
\end{table*}

The closed-loop dynamics between users and intelligent systems have been of growing interest, with several recent studies modeling the feedback loops between users and recommendation systems \cite{rossi2021closed, lanzetti2023impact, sinha2016deconvolving, krauth2025breaking, schmit2018human, haupt2023recommending, deanaccounting}. In this dynamic, the recommendation system curates content based on users' actions, while in turn users' actions, and even their beliefs and opinions, are shaped over time by the recommendations \cite{rossi2021closed, lanzetti2023impact, sprenger2024control, curmei2022towards, jiang2019degenerate, kalimeris2021preference}. This influence is one of the drivers of the negative effects such as extreme opinions, echo chambers, and filter bubbles, as discussed in the introduction. Currently, most proposed solutions to address these issues take the form of the platform moderating and regulating its recommendation algorithm, through mitigation techniques such as diversification~\cite{gharahighehi2023diversification, zheng2021dgcn}, fairness~\cite{biega2018equity, morik2020controlling}, and calibration~\cite{steck2018calibrated, wang2021deconfounded}. Our approach differs from these existing works by exploring whether users \emph{themselves} can adjust their engagement with recommended content to mitigate these issues. 

At the same time, recommendation exposure is not the only driver of opinion formation in many settings. An agent’s opinion is shaped not only by recommendations, but also by social influence through interactions with other agents \cite{sprenger2024control, chandrasekaran2025network, castro2017opinion}. Existing works capture these effects through models of network opinion dynamics in which agents aggregate their neighbors' opinions over time (e.g., DeGroot and Friedkin-Johnsen (FJ) models) \cite{degroot1974reaching, friedkin1990social, proskurnikov2017tutorial}. Our model also includes an FJ model of social influence, and in particular assumes that this effect persists even when a reactive agent chooses not to consume recommended content. That is, we allow recommendation exposure to create indirect effects on opinion drifts through social interactions.

Our proposed model for reactive users builds on \cite{rossi2021closed, lanzetti2023impact, sprenger2024control}, which study how recommendations shape users' opinions over time. Rossi et al. \cite{rossi2021closed} show that interactions between a user and a recommendation system can lead to polarization at the microscopic level. Lanzetti et al. \cite{lanzetti2023impact} study similar opinion dynamics at both microscopic and macroscopic levels, showing that even if each individual's opinion is influenced by recommendations, the overall population's opinion distribution can remain unchanged. Sprenger et al. \cite{sprenger2024control} extend this line of work to a social network setting by combining recommendation dynamics with a variation of the FJ model, and study how recommendation strategies designed to maximize engagement influence opinion formation.

Our work is similar to these papers in how recommendations affect opinions, but differs in two main ways. First, in our model, the recommendation directly affects a user's opinion only when the user consumes content; otherwise, the opinion evolves through an FJ update that captures pure social influence and pulling toward innate opinions. In particular, when an agent clicks, our update aligns with the networked update in \cite{sprenger2024control}. Second, instead of \emph{passive} users as in these prior works, we consider \emph{reactive} users. In \cite{lanzetti2023impact}, users are completely passive and click on \emph{any} recommendations. In \cite{rossi2021closed} and \cite{sprenger2024control}, users decide to click on content based on the similarity between the recommendation and their current preferences, yet their opinions are still influenced by every recommended content, even if they do not click on it. In contrast, our reactive users can intentionally adjust their clicking behavior over time, and experience different opinion dynamics depending on whether they clicked on the recommended content. We show that such adaptive policies can help limit the undesired opinion drifts experienced by passive users in prior works. 

Another related line of work studies strategic users interacting with platforms such as recommendation systems. These include  survey works that find how users adapt their behavior in response to recommendation algorithms~\cite{cen2024measuring}, and theoretical works that model how users strategically decide what content to consume to affect future recommendations~\cite{haupt2023recommending, cen2024user}. Our work differs by explicitly modeling opinion dynamics and social networks.

A preliminary version of this work appeared in \cite{mollabagher2025feedback}, where we first proposed our extended opinion dynamics model for one reactive agent with no social network, and analyzed the short and long-term opinions and utilities of this agent under \emph{deterministic} policies (i.e., consuming content only at pre-determined time steps). Here, we introduce and analyze \emph{stochastic} policies for reactive agents in an extended model with \emph{social influence}, to more realistically allow agents to engage with recommendations probabilistically and socially. Table~\ref{table:comparison} summarizes how our work differs from the related literature.
\section{Model}\label{sec:model}
\subsection{Agent's model} 

We consider a population of $n$ agents connected through a social network. Agent $i\in\{1\dots n\}$ holds an (evolving) \emph{opinion} $x_{i,k}\in [-1,1]$ at each time step $k= 0, 1, 2, \ldots$. We stack the population's opinions in the vector $x_k := [x_{1,k}, \dots, x_{n,k}]^\top\in [-1,1]^n$. This could reflect, for instance, an agent's (political) views on an issue, or preferences for different movie genres. 

At each time step $k$, the platform recommends content $u_{i,k} \in [-1,1]$ to each agent $i$ to consume (e.g., a news article to read, or a movie to watch), collected as $u_{k} := [u_{1,k}, \dots, u_{n,k}]^\top\in [-1,1]^n$. Each agent can decide whether to consume the recommended content, by taking an action $c_{i,k} \in \{0, 1\}$; 
this is a scalar showing whether the agent has clicked on the recommendation (1) or not (0). Let $c_k := \big[c_{1,k},\ldots,c_{n,k}\big]^\top \in \{0,1\}^n$ denote the population's clicking decisions. Let $C_k := \mathrm{diag}(c_k)$.

Through the underlying social network, agents observe and are influenced by the other agents' opinions. We capture this social influence by a row-stochastic matrix $W \in \mathbb{R}_{+}^{n \times n}$, where $w_{ij} \geq 0$ quantifies the relative weight that agent $i$ places on agent $j$'s opinion, with $\sum_{j=1}^n W_{ij}=1$ for all $i$. Then, $(Wx_k)_i = w_i^\top x_k = \sum_{j=1}^n w_{ij} x_{j,k}$ will capture the weighted average of the opinions of agent $i$'s neighbors at time $k$.

Importantly, we let this influence persist even when an agent chooses not to consume content ($c_{i,k}=0$), meaning an agent cannot necessarily fully ``switch off'' external information and may still be exposed to others' opinions through everyday interactions and online social activity. This means that even when agent $i$ does not directly follow a recommendation, the platform may continue to affect its opinion indirectly, since neighboring agents may still consume recommendations, update their opinions accordingly, and propagate this influence through the network to agent $i$.

\textbf{Opinion dynamics model:} Taking the direct effects of the platform recommendation and the effects of social influences together, we propose to model the opinion dynamics of an agent $i\in\{1,\ldots,n\}$ at time $k$ as: 
\begin{equation}\label{eq:user-model-componentwise}
x_{i,k}=
\begin{cases}
\alpha x_{i,0}+\beta w_i^\top x_{k-1}+(1-\alpha-\beta)u_{i,k-1}, & \text{if } c_{i,k-1}=1,\\[4pt]
\frac{\alpha}{\alpha+\beta}x_{i,0}+\frac{\beta}{\alpha+\beta}w_i^\top x_{k-1}, & \text{if } c_{i,k-1}=0.
\end{cases}
\end{equation}

Here, $\alpha \in [0, 1]$ and $\beta \in [0, 1]$ (with $0<\alpha+\beta\leq 1$) are constants determining the relative impact of different factors in shaping the agents' future opinions. We will assume $\alpha\geq\beta$ throughout, under which our model allows the agent to place at least as much weight on the innate opinion as on the social influence. That is, we assume that in absence of influence from the platform, other (internal or external) factors tend to pull the agent back toward its innate opinion, capturing hedonic adaptation~\cite{curmei2022towards}. 

In words, our proposed model in \eqref{eq:user-model-componentwise} states that when agent $i$ clicks on a recommendation, its opinion evolves based on a convex combination of the innate opinion, the social influence of its neighbors' opinions, and the recommendation; when it does not click on the content, \eqref{eq:user-model-componentwise} reduces to a Friedkin-Johnsen (FJ) model \cite{friedkin1990social} with common (homogeneous) stubbornness parameter $\frac{\alpha}{\alpha+\beta}$; the opinion changes based on its innate opinion and social influence. The former (first case in \eqref{eq:user-model-componentwise}) follows the networked update considered in \cite{sprenger2024control}, while the latter (second case) is new in our proposed model, and is introduced due to our consideration of user agency in content consumption. 

When $W=I$ (or $n=1$), the dynamics in \eqref{eq:user-model-componentwise} reduce to decoupled agents, recovering the baseline (no social influence) model:
\begin{equation}\label{eq:user-model-no-network}
    x_{i,k} = 
    \begin{cases} 
    \alpha x_{i,0}+ \beta x_{i, k-1} + (1 - \alpha - \beta) u_{i, k-1}, & \text{if } c_{i,k-1}=1 \\
    \frac{\alpha}{\alpha+\beta} x_{i,0} + \frac{\beta}{\alpha+\beta} x_{i, k-1}, & \text{if } c_{i,k-1}=0
    \end{cases}
\end{equation}
where the former case (under $c_{i,k-1}=1$) is the same as the opinion dynamics considered in previous work \cite{rossi2021closed, lanzetti2023impact}, while the latter case (under $c_{i,k-1}=0$) is again new to our model due to the consideration of reactive users.

\textbf{Vector and compact form of the opinion dynamics model:} Using \eqref{eq:user-model-componentwise}, the vector of all agents' opinions evolves as follows:
\begin{equation}\label{eq:user-model}
\begin{aligned}
x_k=\;& C_{k-1}\Big(\alpha x_0+\beta W x_{k-1}+(1-\alpha-\beta)u_{k-1}\Big) \\
&\;+\big(I-C_{k-1}\big)\Big(\tfrac{\alpha}{\alpha+\beta}x_0+\tfrac{\beta}{\alpha+\beta}W x_{k-1}\Big).
\end{aligned}
\end{equation}
Here, $x_0\in [-1,1]^n$ is the population's initial or \emph{innate} opinions vector. We can equivalently express our proposed model of opinion dynamics \eqref{eq:user-model} in a compact form. Let $\zeta:=\alpha+\beta$ and $b:=\frac{\beta}{\alpha+\beta}$, and recall $C_{k-1}=\mathrm{diag}(c_{k-1})$. Let
\begin{equation}\label{eq:PQ-def}
Q_{k-1}:=bI+(\beta-b)C_{k-1},\qquad 
P_{k-1}:=(1-\zeta)C_{k-1}.
\end{equation}
Then \eqref{eq:user-model} can be written as
\begin{equation}\label{eq:user-model-PQ}
x_k \;=\; (I-Q_{k-1}-P_{k-1})x_0 \;+\; Q_{k-1}W x_{k-1} \;+\; P_{k-1}u_{k-1}.
\end{equation}

Intuitively, $Q_{k-1}$ is a diagonal matrix that captures how strongly each agent carries over the \emph{social influence term} $W x_{k-1}$ to the next step, with different weights depending on whether the agent clicked or not. In contrast, $P_{k-1}$ is a diagonal matrix that captures the \emph{direct pull of the recommendation} on the next-time step opinion for agents who click at time $k-1$ (and it is zero for non-clickers). We will use this compact form throughout our analysis.

\textbf{Agents' utility:} Clicking on a recommended content has two consequences: it influences the evolution of the agent's opinion (as detailed above), and it generates a benefit/reward for the agent (due to its entertainment value, providing knowledge, etc.). To model the latter, we define an \emph{agent reward function} $R^A: \mathbb{R}_{\geq 0} \to \mathbb{R}_{\geq 0}$, which determines the instantaneous reward for agent $i$ at each time $k$, evaluated at $|x_{i,k} - u_{i,k}|$ (the distance between agent $i$'s current opinion and the consumed content). We assume $R^A$ is a non-increasing function: rewards are higher when the the difference between the recommendation and the opinion is smaller (e.g., a recommended article is close to the agent's views). 

Accordingly, we let the agent $i$'s utility for a horizon $K$ be:
\begin{equation}\label{eq:agent-utility}
    U_i(h_{i,K} ^A) = \lambda \frac {1} {K} \sum_{k=0}^{K-1} c_{i,k} R^A (|x_{i,k} - u_{i,k}|) -  (1 - \lambda) |x_{i,K} - x_{i,0}|~.
\end{equation}

Here, $h^A_{i,K}$ denotes the agent $i$'s history up to time $K$, which contains the set of all its opinions $\{x_{i,k}\}_{k \in [0:K]}$, the platform's recommendations to agent $i$ $\{u_{i,k}\}_{k \in [0:K-1]}$, agent $i$'s actions $\{c_{i,k}\}_{k \in [0:K-1]}$, and agent $i$'s rewards $\{R^A(|x_{i,k} - u_{i,k}|)\}_{k \in [0:K-1]}$. 
In words, the utility states that the agent wants to maintain an opinion close to its initial opinion after $K$ time steps, while also deriving benefit from consuming content during this time, with $\lambda\in[0,1]$ capturing the relative importance of each goal.

\subsection{Platform's model} 
The objective of the platform is to recommend contents that agents choose to consume. At each time step $k$, if agent $i$ clicks on the recommended content ($c_{i,k}=1$), the platform receives a reward of $1$, and $0$ otherwise. Accordingly, we let the platform's payoff from its interactions with agents over a horizon $K$ be the average click-through rate (CTR) across the population:
\begin{equation}\label{eq:platform-utility}
    \Pi(h_K^P)\;=\;\frac{1}{nK}\sum_{k=0}^{K-1}\sum_{i=1}^n c_{i,k},
\end{equation}
where $h_{K}^P$ denotes the platform's history up to time $K$, consisting of the set of agents' content consumption actions $\{c_{i,k}\}_{i\in[n],\,k\in[0:K-1]}$.
\section{Agent's and Platform's Policies}\label{sec:policy}

\subsection{Agent's Policies}

On the agents' side, we consider two stochastic policies in both our analysis and numerical experiments: ``\emph{fixed}'' and ``\emph{adaptive decreasing}'' clicking policies. 
In both policies, an agent $i$ clicks on the content with some probability $\gamma_{i,k}$ at each time step $k$. The policies differ in how $\gamma_{i,k}$ is selected.

In the \emph{fixed} policy (Algorithm~\ref{alg:policy-one}), $\gamma_{i,k} \in [0, 1]$ is kept constant over time. If we let $\gamma_{i,0}=1$, this will be equivalent to the ``always click'' policy of \emph{passive agents} considered in prior works.

\begin{algorithm}[t]
    \hspace*{\algorithmicindent} \textbf{Input:} Total time steps $K$; constant $\gamma_{i,0} \in [0, 1]$.
    \caption{Agent $i$'s ``Fixed'' Clicking Policy}\label{alg:policy-one}
    \begin{algorithmic}
        \For{$k \in  \{0, 1, 2, \dots, K-1\}$}
            \State At time step $k$, click with probability $\gamma_{i,0}$.
        \EndFor
    \end{algorithmic}
\end{algorithm}

In the \emph{adaptive decreasing} policy (Algorithm~\ref{alg:policy-two}), agent $i$ starts with an initial clicking probability $\gamma_{i,0}\in[0,1]$. After each opinion update, if the agent finds that its next opinion has deviated substantially from its innate opinion (if $|x_{i,k+1}-x_{i,0}|\geq \delta$ for a given tolerance $\delta>0$), then the agent reduces its clicking probability for the next time step to $\gamma_{i,k+1}=\frac{\gamma_{i,k}}{\kappa}$, where $\kappa\geq 1$ is a reduction rate; otherwise, the click rate will remain unchanged at $\gamma_{i,k+1}=\gamma_{i,k}$. This is a \emph{reactive agent} policy.

\begin{algorithm}[t]
    \hspace*{\algorithmicindent} \textbf{Input:} Total time steps $K$; initial clicking probability $\gamma_{i,0} \in [0, 1]$; clicking decrease rate $\kappa\geq 1$; drift tolerance $\delta>0$.  
    \caption{Agent $i$'s ``Adaptive Decreasing'' Clicking Policy}\label{alg:policy-two}
    \begin{algorithmic}
        \For{$k \in  \{0, 1, 2, \dots, K-1\}$}
            \State At time step $k$, click with probability $\gamma_{i,k}$.
            \If{$|x_{i,k+1} - x_{i,0}| \geq \delta$}
                \State Reduce clicking probability to $\gamma_{i, k+1} \gets \frac{\gamma_{i,k}}{\kappa}$.
            \Else
                 \State Keep clicking probability at $\gamma_{i, k+1} \gets \gamma_{i,k}$.
            \EndIf
        \EndFor
    \end{algorithmic}
\end{algorithm}

\subsection{Platform's Policies}\label{sec:platform-policy}
We consider two platform policies: \emph{``fixed recommendation''} for analytical results, and \emph{``explore periodically''} for numerical experiments. At each time $k$, the platform selects a recommendation vector $u_k\in[-1,1]^n$, where $u_{i,k}$ denotes the scalar recommendation shown to agent $i$.

In the ``fixed recommendation'' policy, the platform makes the same scalar recommendation $u \in [-1,1]$ to all agents at all times, i.e., $u_{i,k} = u$ for all $i\in[n],\ k\ge 0$. This allows us to contrast our results with that of \cite{lanzetti2023impact}, which has analyzed the impacts of this recommendation policy on \emph{passive} agents.

In the \emph{``explore periodically''} policy, a variation of the policy considered in prior works \cite{lanzetti2023impact, rossi2021closed}, the platform always chooses the action which leads to the highest click rate observed so far, with the exception of some exploration steps, in which the recommendation will be chosen from a probability distribution; 
i.e., $u_k = u \mathbf{1}_n$ where $u \sim \rho$ on $[-1,1]$ for $k \in \{0, T, 2T, \ldots\}, \text{ and } T\in\mathbb{N}$. This policy resembles $\epsilon$-greedy exploration in the multi-armed bandit/reinforcement learning literature. The recommendation shown to all agents at exploitation steps is selected among previously recommended values, such that the one with the highest empirical CTR so far is selected. Specifically, for any scalar candidate $u$ that has been recommended before time $k$, we define
\[
{\mathrm{CTR}}_k(u) =\frac{\sum_{t=0}^{k-1}\mathbf{1}\{u_t=u\}\sum_{i=1}^n c_{i,t}}
{n\sum_{t=0}^{k-1}\mathbf{1}\{u_t=u\}},
\]
whenever the denominator is nonzero. Then, at exploitation time steps, the platform selects
\begin{align}\label{eq:platform-recommendation}
    u_k \in \operatorname*{arg\,max}_{u\in\{u_0,\ldots,u_{k-1}\}}{\mathrm{CTR}}_k(u)~,
\end{align}
with ties broken at random. If the platform has not observed any clicks yet, it defaults to exploration.

\section{Analysis}\label{sec:analysis}

In this section, we analytically characterize the evolution of agents' opinions, both \emph{ex-post} (after a realized click sequences) and \emph{ex-ante} (the expectation over clicks) as well as their utilities, in finite time and in the limit, when agents follow one of the policies outlined in Section~\ref{sec:policy}. Our main Propositions are followed by proof sketches; the full proofs for all results are given in the Appendix.

\subsection{Ex-post opinions $x_k$} Recall that at each time step $k$, each agent $i$ clicks on the recommendation with some (policy-dependent) probability $\gamma_{i,k} \in [0,1]$, making each click event $c_{i,k}$ a binary random variable $c_{i,k} \sim \mathrm{Bernoulli}(\gamma_{i,k})$. First, we consider a realization of this sequence of random variables for each agent $i$ and characterize the agents' ex-post opinion $x_k$ under it.

\begin{lemma}\label{lem:convex-combination-opinion}
    Assume agents follow the click realizations $c_0, \ldots, c_{k-1}$. Then, the agents' ex-post opinion vector $x_{k}$ at time  step $k$ is given by
    \[x_k = M_k x_0 + G_k u_0\]
    where 
    \begin{align*}
        M_k &= \Phi(k,0)+\sum_{t=0}^{k-1} \Phi(k,t+1)\,\big(I-Q_t-P_t\big),\\
        G_k &=\sum_{t=0}^{k-1} \Phi(k,t+1)P_t.
    \end{align*}
    with $P_t$ and $Q_t$ defined in~\eqref{eq:PQ-def}, and $\Phi(k,t+1):=\prod_{\tau=t+1}^{k-1}(Q_{\tau} W), ~ \Phi(k,k):=I$. 
    Moreover, $M_k$ and $G_k$ are entrywise nonnegative and satisfy $(M_k+G_k)\mathbf{1}=\mathbf{1}$. 
\end{lemma}

Lemma~\ref{lem:convex-combination-opinion} shows that the opinion $x_{i,k}$ of each agent $i$ is a convex combination of the innate opinions $\{x_{j,0}\}_{j=1}^n$ and the common recommendation $u$.
The actual weights in this combination depend on the realized click sequences, and therefore on the agents' adopted policies, as further detailed below.

First, based on the compact form of the opinion dynamics \eqref{eq:user-model-PQ}, we note that the recommendation enters the one-step update only through the term $P_t u_0$ with $P_t=(1-\zeta)C_t$. Due to this, 
policies that induce more click events increase the cumulative contribution of $\{P_tu_0\}_{t=0}^{k-1}$ to $x_k$ (directly for clicking agents and indirectly through propagation via the social influence term $W x_t$), and therefore tend to move the population opinion toward $u_0$, as intuitively expected. Next, we assess the impacts of $\alpha$ (the weight of the innate opinions) and $\beta$ (the weight of the neighbors' opinions) on $x_k$. If $\alpha$ increases, the pull toward innate opinions increases, both when clicking (since the coefficient on $x_0$ is $\alpha$) and when not clicking (since $1-b=\frac{\alpha}{\alpha+\beta}$ increases), therefore reducing the overall influence of $u_0$ on $x_k$. The case for $\beta$ is more complex, and increasing it does not guarantee an increase or decrease in the impact of the recommendation $u_0$ on $x_k$. A higher $\beta$ increases the persistence of past opinions through the social influence term $Wx_{t}$ (so the influence of past clicks diminish more slowly), while at the same time reducing the direct influence of each individual click through the factor $1-\zeta$ in $P_t$. Therefore, one effect may dominate the other based on the exact realizations of clicks and the structure of the network.

\subsection{Ex-ante expected opinions $\mathbb{E}[X_k]$}  
We next note that different agent policies permit the same possible click sequences, but with different probability for each trajectory. Therefore, to compare opinion evolution under different policies, we leverage the results of Lemma~\ref{lem:convex-combination-opinion}, and compare the ex-ante expected value of the population's opinion $\mathbb{E}[X_k]$, where $X_k$ is the population's opinion at time step $k$, a random vector dependent on the random click variables $c_{i,t} \sim \mathrm{Bernoulli}(\gamma_{i,t}).$

We will start with the benchmark setting in which all agents follow the fixed clicking policy with a common rate $\gamma_0$; we denote the resulting expected opinion by $\mathbb{E}[X_k^\text{all}]$. After that, to isolate the role of a single strategic agent $s$, we consider a mixed population in which all agents $i\neq s$ are passive and click at every time step (i.e., follow the fixed clicking policy with $\gamma_{i,0}=1$), while agent $s$ is reactive and follows one of two policies: fixed vs. adaptive decreasing. We write $X_k^{(\texttt{p})}$ for the resulting opinion process under this setting, with $\texttt{p}=1$ corresponding to agent $s$ using the fixed clicking policy with rate $\gamma_{s,0}$ and $\texttt{p}=2$ corresponding to agent $s$ using the adaptive decreasing policy.

\textbf{Benchmark: all passive agents}
We begin with the expected opinion of \emph{passive} agents under the \emph{fixed} clicking policy (policy~\ref{alg:policy-one}). 

\begin{proposition}\label{prop:expectation-opinion-fixed}
     Let $p := (1-\zeta)\gamma_0$ and $q := b + (\beta-b)\gamma_0$. Then, the expected opinion of the population at the time step $k$ under the fixed policy~\ref{alg:policy-one} is given by: \[\mathbb{E}[X_k^\text{all}] = \big((qW)^k + (1-q)A_k\big)x_0 + pA_k\,(u_0-x_0)\]
    where $A_k:=(I-qW)^{-1}\big(I-(qW)^k\big)$.
\end{proposition}

\emph{Proof sketch.} We start from the one-step opinion dynamics \eqref{eq:user-model-PQ} and take expectations. Under the \emph{fixed clicking} policy, clicks are i.i.d. Bernoulli$(\gamma_0)$ across agents and time, which implies $\mathbb{E}[Q_k]=qI$ and $\mathbb{E}[P_k]=pI$.
Moreover, $Q_k$ depends only on the clicks at time $k$, while $X_k$ depends only on past clicks up to time $k-1$; hence $Q_k$ is independent of $X_k$. This yields a closed linear recursion for $\mathbb{E}[X_k^\text{all}]$ driven by the network matrix $W$. Unrolling the recursion gives a matrix geometric sum; we write it in closed form by showing $I-qW$ is invertible (using a null-space argument and the fact that $q<1$), and then applying the matrix geometric-series identity.
\hfill\qedsymbol

Here, $A_k=\sum_{t=0}^{k-1}(qW)^t$ aggregates the effect of social mixing over paths of length up to $k-1$ (self, neighbors, neighbors-of-neighbors, etc.), with longer paths discounted by $q^t$. Using this fact, Proposition~\ref{prop:expectation-opinion-fixed} shows that the influence of the recommendation $u_0$ on the expected opinion increases over time under the fixed clicking policy of passive agents through the term $pA_k(u_0-x_0)$, and since $A_{k+1}=A_k+(qW)^k\succeq A_k$, the corresponding gain grows with $k$ (and eventually saturates as $k$ becomes large because $q<1$). Intuitively, this is because as time goes on, the recommendation can affect the population both directly (through agents who click) and indirectly (as opinions are repeatedly mixed through the network), so its overall effect builds up over time.

To connect with the model in which there is no social influence (the model studied in prior work), we consider a special case of Proposition~\ref{prop:expectation-opinion-fixed} with $W=I$, under which agents are decoupled and the expected opinion admits a scalar closed form.

\begin{corollary}\label{cor:expectation-opinion-fixed-no-social}
    Consider the setting of Proposition~\ref{prop:expectation-opinion-fixed}. If $W=I$, then under the fixed policy~\ref{alg:policy-one},
    \[\mathbb{E}[X_k^\text{all}] = x_0 + p\,\frac{1-q^k}{1-q}\,(u_0-x_0).\]
\end{corollary}

Corollary~\ref{cor:expectation-opinion-fixed-no-social} shows that, in the absence of social influence ($W=I$), the expected opinion is a convex combination of $x_0$ and $u_0$. In the special case $\gamma_0 = 1$, the result of Corollary~\ref{cor:expectation-opinion-fixed-no-social} takes the same form as the deterministic dynamics in \cite{lanzetti2023impact}, that is, recovering the benchmark of no social influence model, where the initial opinion and the recommendation are fixed at $x_0$ and $u_0$, respectively.

Proposition~\ref{prop:expectation-opinion-fixed} also allows us to characterize the long term expected opinion under the fixed policy.

\begin{corollary}\label{cor:expectation-opinion-fixed-infinity}
    Consider the settings of Proposition~\ref{prop:expectation-opinion-fixed}. Then, $$\lim_{K\to\infty}\mathbb{E}[X_K^\text{all}] = (I-qW)^{-1}\Big((1-q)x_0 + p\,(u_0-x_0)\Big).$$
\end{corollary}

Corollary~\ref{cor:expectation-opinion-fixed-infinity} shows that under a fixed recommendation $u_0$ and the fixed policy, the population's expected opinion converges to a unique steady state. In the steady state, $(I-qW)^{-1}$ captures how social influence repeatedly propagates through the network over time, as is standard in the FJ social influence models \cite{proskurnikov2017tutorial, sprenger2024control}. 

In the special case $\gamma_0=1$, again the dynamics are deterministic, so the expected opinion coincides with the opinion itself. The steady state opinion becomes $(I-\beta W)^{-1}\Big(\alpha x_0 + (1-\alpha-\beta)u_0\Big)$, which is consistent with the setting of fixed recommendations in \cite{sprenger2024control}.

\textbf{Mixed setting: one deviating agent}
To compare the long-run behavior of a \emph{passive} vs.\ \emph{reactive} strategic agent $s$, we now turn to a mixed population in which all non-strategic agents follow the fixed clicking policy with the fixed rate of $\gamma_{i,0} = 1$, while a single reactive agent follows either the \emph{fixed} (policy~\ref{alg:policy-one}) or \emph{adaptive decreasing} (policy~\ref{alg:policy-two}) clicking policy.

We begin by characterizing the long-run expected opinions when one deviating agent adopts a fixed clicking policy with a different rate than others in the population. We do so under the following assumption.  

\begin{assumption}\label{ass:no-outgoing-s}
Agent $i$ is influenced by, but does not influence, other agents; that is, $w_{is}=0$ for all $i\neq s$. 
\end{assumption}

Assumption~\ref{ass:no-outgoing-s} is mild in a large-population dense network, in which the influence of a single agent's opinion on everyone else is negligible. In our analysis, this means that we consider the first-order effect of reduced content consumption of the strategic agent adopting a policy different from others, but ignore the second-order effects of the strategic agent's deviation, as this agent's evolving opinion does not enter other agents' social signals. With this, we can characterize agents' long-term opinions in this mixed setting. 

\begin{corollary}\label{cor:expectation-opinion-mixed-fixed-infinity}
Consider the settings of Corollary~\ref{cor:expectation-opinion-fixed-infinity} under Assumption~\ref{ass:no-outgoing-s}, with $\gamma_{i,t}=1$, for $i \neq s, \forall t$. Let $p:=(1-\zeta)\gamma_0$ and $q:=b+(\beta-b)\gamma_0$. Then,
\[
\lim_{K\to\infty}\mathbb{E}\!\left[X_K^{(1)}\right]
=
\begin{bmatrix}
x_{-s}^\star\\[1mm]
x_{s,\infty}^{(1)}
\end{bmatrix},
\]
where
\begin{align*}
    x_{-s}^\star &= (I-\beta W_{-s,-s})^{-1}\Big(\alpha x_{-s,0}+(1-\zeta)u\mathbf 1_{-s}\Big),\\
    x_{s,\infty}^{(1)} &= (1-q\,w_{ss})^{-1}
    \Big((1-q-p)\,x_{s,0}+p\,u+q\,w_{s,-s}\,x_{-s}^\star\Big).
\end{align*}
\end{corollary}

Corollary~\ref{cor:expectation-opinion-mixed-fixed-infinity} characterizes the long-run expected opinion in the mixed population when the strategic agent $s$ remains \emph{passive} but clicks with a reduced constant rate $\gamma_0$. Under Assumption~\ref{ass:no-outgoing-s}, agents $i\neq s$ converge to the same limit $x_{-s}^\star$ as in the all click dynamics, while the long-run expected opinion of agent $s$ is shifted toward the recommendation through the $(p,q)$ terms. 

We now turn to the \emph{reactive} case and show that under the adaptive decreasing policy, the reactive agent's clicking rate decreases over time and eventually vanishes almost surely.

\begin{proposition}\label{prop:gamma-convergence}
    Assume that agent $s$ is the only reactive agent and follows the adaptive decreasing policy~\ref{alg:policy-two} with $\kappa>1$. Let \(\gamma_{i,0}=1\) for all agents \(i\).
    Let $\alpha, \beta, \delta$ be such that under the all click dynamics, agent $s$'s opinion is pushed at least $\delta$ away from its innate opinion (formally, $\left|e_s^\top\Big((I-\beta W)^{-1}(\alpha x_0+(1-\zeta)u_0)-x_0\Big)\right| > \delta$). Then, under Assumption~\ref{ass:no-outgoing-s}, the reactive agent's clicking probability converges to zero almost surely:
    \[
    \mathbb{P}\left( \lim_{k \to \infty} \gamma_{s,k} = 0 \right) = 1.
    \]
\end{proposition}

\emph{Proof sketch.} Two complementary proofs are provided for this finding: the first is based on the \emph{Borel-Cantelli Lemma} (using a block argument), and the second is based on the structure of the opinion dynamics and the opinion process. (The second approach draws interesting connections between our model and certain stochastic models used in finance.)

The first approach relies on a block construction and the Borel–Cantelli Lemma. The main idea is to define a region in which the threshold for decreasing the clicking probability is not triggered, and then show that the deviation counter cannot stop increasing. Using the all click dynamics for all the population including $s$, we show that if agent $s$ clicks for a sufficiently long sequence of \emph{consecutive} time steps, then its opinion must cross the tolerance threshold $\delta$. Now suppose, for contradiction, that the clicking probability eventually stops decreasing and stays bounded away
from zero. In that case, the events ``agent $s$ clicks for the next $M$ consecutive steps'' have a fixed positive probability and occur independently across disjoint time blocks; by the Borel-Cantelli lemma, such events occur infinitely often almost surely. Each such event forces a threshold crossing, which contradicts the assumption that no further reductions happen. Therefore threshold crossings occur infinitely often, so the clicking probability is reduced infinitely often, and hence $\gamma_{s,k}\to 0$ almost surely.

The main idea behind the second approach is to track the time steps at which the clicking probability remains bounded away from zero. Each time the process enters to the tolerance region (where no reduction is triggered) while the clicking probability is still relatively large, there is a positive probability that agent $s$ clicks for a sufficiently long stretch of consecutive time steps. As shown earlier, any such drift necessarily pushes the opinion outside the tolerance threshold, which triggers a reduction in the clicking probability. Importantly, as long as the clicking probability stays above a fixed level, this probability is bounded below by the same constant whenever we are in this situation. This reasoning is similar to the \emph{positive crossing (first-passage) probability} argument used in financial models~\cite{bakshi2010first}. Inspired by this idea, we show that the probability that the process never triggers such a threshold crossing across infinitely many such time steps is zero. Consequently, the clicking probability cannot remain bounded away from zero infinitely often, and therefore \( \gamma_{s,k} \to 0 \) almost surely.
\hfill\qedsymbol

In words, Proposition~\ref{prop:gamma-convergence} shows that the decrease in the clicking probability will almost surely get triggered infinitely often, so that agent $s$ will ultimately stop consuming content. Accordingly, we can characterize the agents' long-run opinion.

\begin{corollary}\label{cor:expectation-opinion-adaptive-infinity}
    Consider the settings of Proposition~\ref{prop:gamma-convergence} under Assumption~\ref{ass:no-outgoing-s}. Then,
    \[\lim_{K\to\infty}\mathbb{E}\!\left[X_K^{(2)}\right]
    =
    \begin{bmatrix}
    x_{-s}^\star\\[1mm]
    x_{s,\infty}^{(2)}
    \end{bmatrix},
    \]
    where
    \begin{align*}
        x_{-s}^\star &= (I-\beta W_{-s,-s})^{-1}\Big(\alpha x_{-s,0}+(1-\zeta)u\mathbf 1_{-s}\Big)~,\\
        x_{s,\infty}^{(2)} &= (1-bw_{ss})^{-1}\Big((1-b)x_{s,0}+b\,w_{s,-s}\,x_{-s}^\star\Big)~.
    \end{align*}
\end{corollary}

Based on Proposition~\ref{prop:gamma-convergence}, the clicking probability of agent $s$ converges to zero almost surely, meaning that agent $s$ eventually stops clicking. From that time on, agent $s$ is no longer directly affected by the recommendation, and its opinion evolves only through its innate opinion and social influence. At the same time, all agents $i\neq s$ continue to click and therefore keep following the all click update. Under Assumption~\ref{ass:no-outgoing-s} (i.e. $w_{is}=0$ for all $i\neq s$), the evolution of the subpopulation $-s$ is
independent of $x_{s,K}$ for any time. Therefore, the opinions of agents $i\neq s$ converge to a steady state $x_{-s}^\star$ (the same steady state
as if agent $s$ did not exist). Given this limit, agent $s$ is pulled toward a fixed point $x_{s,\infty}^{(2)}$, which can be interpreted as a convex combination of its innate opinion $x_{s,0}$ and the neighbors' steady-state opinion vector $w_{s,-s}x_{-s}^\star$. As a result, the expected value of $X_K^{(2)}$ converges to $\begin{bmatrix}
    x_{-s}^\star\\[1mm]
    x_{s,\infty}^{(2)}
    \end{bmatrix}$ almost surely.

To compare the long-run behavior of a \emph{passive} vs. \emph{reactive} strategic agent $s$ in the same mixed-population setting, we can contrast Corollary~\ref{cor:expectation-opinion-mixed-fixed-infinity} (passive $s$ with fixed clicking rate $\gamma_0$) and Corollary~\ref{cor:expectation-opinion-adaptive-infinity} (reactive $s$ with adaptive decreasing policy). Under Assumption~\ref{ass:no-outgoing-s}, the long-run behavior of agents $i\neq s$ is identical in both cases and converges to $x_{-s}^\star$. The distinction is in agent $s$: if $s$ stays passive, it may keep consuming platform content, so the recommendation continues to enter its expected update, and its long-run expected opinion converges to $x_{s,\infty}^{(1)}$; if $s$ is reactive, it eventually stops clicking, so the platform no longer affects $s$ directly in the long run, so its limit becomes $x_{s,\infty}^{(2)}$.

In the special case $W=I$ (no social influence), this comparison becomes even more clear. Under the fixed clicking policy, Corollary~\ref{cor:expectation-opinion-fixed-infinity} reduces to the scalar limit $\lim_{K\to\infty}\mathbb{E}[X_{s,K}^{(1)}]
= x_{s,0}+\frac{p}{1-q}(u-x_{s,0})$, while under the adaptive decreasing policy $\lim_{K\to\infty}X_{s,K}^{(2)}=x_{s,0}$ almost surely (and hence also in expectation). In other words, under the \emph{fixed} clicking policy, the expected long-run opinion of a passive agent stays further away from the innate opinion, whereas under the \emph{adaptive decreasing} policy, a reactive agent can prevent such opinion drifts. 

\subsection{Agents' expected utilities} Our analysis so far shows that the reactive agent can prevent opinion drifts in the long-run. However, this comes at the expense of reduced content consumption, as shown in Proposition~\ref{prop:gamma-convergence}. To account this trade-off, we next compare the agent's utility under the two policies. We start with agents' expected utility in the long-run.

We compare the long-run expected utilities under the following assumption on the reactive agent $s$, which guarantees that the deviation $|X_{s,k}-x_{s,0}|$ for a fixed agent $s$ keeps a fixed sign, so we can simplify the drift term in the utility. 

\begin{assumption}\label{ass:sign-preservation}
For the reactive agent $s$, the recommendation and the innate opinions of all agents influencing $s$ lie on the same side of $x_{s,0}$, i.e., either
\begin{align*}
x_{s,0}&\le u
\quad\text{and}\quad
x_{s,0} \leq x_{j,0}\ \quad \forall j\text{ such that } w_{sj}>0, \quad \text{or }\\
x_{s,0}&\ge u
\quad\text{and}\quad
x_{s,0} \geq x_{j,0}\ \quad \forall j\text{ such that } w_{sj}>0.
\end{align*}
\end{assumption}

Under this assumption, we can derive closed-form expressions for the agents' long-run expected utilities under the two policies.

\begin{lemma}\label{lem:compare-utility-infinite}
    Consider the mixed-population setting of Corollary~\ref{cor:expectation-opinion-mixed-fixed-infinity} (passive $s$ with fixed rate $\gamma_0$) and Corollary~\ref{cor:expectation-opinion-adaptive-infinity} (reactive $s$ with adaptive decreasing policy). Suppose Assumption~\ref{ass:no-outgoing-s} and Assumption~\ref{ass:sign-preservation} hold for agent $s$ and assume $R^A(|x_{i,k}-u_{i,k}|)=1$ for all $i,k$. Then, under the fixed clicking policy,
    \[\lim_{K\to\infty}\mathbb{E}\!\left[U_s^{(1)}(h_{s,K}^{(A)})\right]=\lambda\,\gamma_0-(1-\lambda)\,\sigma_s\Big(x_{s,\infty}^{(1)}-x_{s,0}\Big),\]
    
    and under the adaptive decreasing policy,
    \[\lim_{K\to\infty}\mathbb{E}\!\left[U_s^{(2)}(h_{s,K}^{(A)})\right]=-(1-\lambda)\,\sigma_s\Big(x_{s,\infty}^{(2)}-x_{s,0}\Big),
    \]
    
    where \(\sigma_s := \operatorname{sign}(u-x_{s,0})\), and $x_{s,\infty}^{(1)}$ and $x_{s,\infty}^{(2)}$ are given in Corollary~\ref{cor:expectation-opinion-mixed-fixed-infinity} and Corollary~\ref{cor:expectation-opinion-adaptive-infinity}, respectively.
\end{lemma}

Lemma~\ref{lem:compare-utility-infinite} can be interpreted as follows. Under the fixed policy, the agent keeps receiving the consumption benefit $\lambda\gamma_0$ but also accumulates a persistent drift away from $x_{s,0}$. Under the adaptive decreasing policy, the agent eventually stops clicking, so the long-run consumption term vanishes, but the long-run drift is smaller since the recommendation no longer enters the update directly. If content consumption matters more to the agent (formally, if
\[
\lambda > \frac{1}{\,1+\dfrac{1- q w_{ss}}{\sigma_s\,\Delta_s}\,},\] where $\Delta_s~:=~
(1-\zeta)[(u-x_{s,0})-\; b\,\frac{w_{s,-s}(x_{-s}^\star-x_{s,0}\mathbf 1_{-s})}{1-bw_{ss}}]$), then the expected utility under the passive {fixed} clicking policy becomes positive and can outperform that of the {adaptive decreasing} clicking policy. On the flip side, for agents who value maintaining their innate opinion, the {adaptive decreasing} clicking policy yields higher long-run expected utility and will therefore be preferred by agents.

While this establishes that there are scenarios in which the reactive policy is overall preferred to a passive policy, the result of Lemma~\ref{lem:compare-utility-infinite} is in the limit, and may seem to imply that policy~\ref{alg:policy-two} is only preferred because it leads to full disengagement from the platform in the long-run. However, we next contrast the policies' expected utilities in finite time horizons, and show that there are similar conditions (namely, sufficiently small $\lambda$) under which a reactive agent following the adaptive policy~\ref{alg:policy-two} can outperform a passive one following the fixed policy~\ref{alg:policy-one} in finite time.

\begin{proposition}\label{prop:finite-adaptive-better-than-fixed}
Consider the mixed-population setting of Lemma~\ref{lem:compare-utility-infinite} over a finite horizon of length $K$, under Assumption~\ref{ass:no-outgoing-s} and ~\ref{ass:sign-preservation}. Suppose that the adaptive policy~\ref{alg:policy-two} triggers at least one reduction by time \(K\). Let \(x^\star=\big[(x_{-s}^\star)^\top,\ x_s^\star\big]^\top\) denote the steady state of the all click dynamics (i.e., when every agent clicks at every step), as characterized in Corollary~\ref{cor:expectation-opinion-fixed-infinity}. If 
\[
\lambda \;<\; \frac{1}{\,1+\dfrac{K-M-\frac{1-\kappa^{-(K-M)}}{\kappa-1}}{K(1-\kappa^{-1})\,G_M}\,},
\]
then the adaptive policy~\ref{alg:policy-two} is guaranteed to outperform the fixed policy~\ref{alg:policy-one} in expectation.
Here,
\begin{align*}
M&:=\min\Big\{m\in\mathbb{N}:\ \beta^m(1+\|x^\star\|_\infty)\le |x_s^\star-x_{s,0}|-\delta\Big\}~,\\
    G_M&:=\sum_{k=M}^{K-1}\rho^{\,K-1-k}g_k,\\
    \rho&:=(b+(\beta-b)\gamma_0)w_{ss},\\
    g_k &:= \sigma_s(1 - \zeta)\Big((u-x_{s,0}) -b(w_{s,-s}X_{-s,k}-x_{s,0} + w_{ss}) \Big),
\end{align*}
where $\sigma_s := \operatorname{sign}(u-x_{s,0})$, $\gamma_0$ is the fixed clicking probability of agent $s$ under policy~\ref{alg:policy-one}, and \(\{X_{-s,k}\}_{k\ge 0}\) is deterministic and given by Proposition~\ref{prop:expectation-opinion-fixed} applied to subnetwork $(W_{-s,-s}, x_{-s,0})$ with initial clicking probability of 1.
\end{proposition}

\emph{Proof sketch.}
Under Assumption~\ref{ass:no-outgoing-s}, all agents $i\neq s$ always click, so their opinions evolve deterministically; this lets us treat the neighbors’ contribution to agent $s$'s opinion as a known (deterministic) signal over time. We compare the expected utilities under the fixed and adaptive policies and split the gap into two pieces: a \emph{clicking gap} term (adaptive clicks less on average) and a \emph{drift gap} term (adaptive keeps the opinion closer to $x_{s,0}$). By Assumption~\ref{ass:sign-preservation}, we can drop the absolute value in the drift term by fixing the sign $\sigma_s$ and work with signed quantities. For the clicking gap term, we use the definition of $M$: before time $M$ no reduction can happen, and after that the adaptive policy can only reduce by factors of $\kappa$, so the click probability is bounded below by the most conservative evolution $\frac{\gamma_0}{\kappa^j}$. Averaging this lower bound gives an upper bound on the expected clicking probability ($\gamma_0-\bar\gamma_K$). For the drift gap term, we write one step recursions for the opinion means under the fixed and adaptive policies and subtract them to obtain a recursion for the signed mean gap $d_k$. The only subtle term is the part where the random click probability $\gamma_{s,k}$ multiplies the current opinion; we bound this term using $|X^{(2)}_{s,k}|\le 1$ and $\gamma_{s,k}\le \gamma_0$ after reductions. This yields a clean lower bound of the form $d_{k+1} \geq \rho d_k + g_k\Delta_k$, 
which can be iterated to lower bound $d_K$ in terms of $G_M$. Combining the clicking and drift gap bounds and rearranging gives the stated sufficient condition on $\lambda$. 
\hfill\qedsymbol

Intuitively, when there is no deviation under the adaptive policy~\ref{alg:policy-two}, the clicking probability never decreases, so the expected utility under fixed and adaptive decreasing policies is the same. If a deviation is triggered, the adaptive policy becomes more conservative from that point on (it reduces future clicking), which costs some expected clicks but also reduces how far the opinion drifts away from $x_{s,0}$. Ultimately, Proposition~\ref{prop:finite-adaptive-better-than-fixed} states that the adaptive decreasing policy wins in expected utility if drift reduction dominates the click loss. Note also that the bound on $\lambda$ is a sufficient but not necessary condition; outside this range, the adaptive policy~\ref{alg:policy-two} may still outperform the fixed policy~\ref{alg:policy-one}.
\section{Numerical Experiments}\label{sec:numerical}

In this section, we use numerical experiments to validate our theoretical findings of Section~\ref{sec:analysis} for the \emph{fixed recommendation} policy of the platform, and also provide numerical results for the \emph{explore periodically} policy of the platform. Throughout, we consider the reward functions $R^A (|x_{i,k} - u_{i,k}|) = 1 - d|x_{i,k} - u_{i,k}| $, for any agent $i$, where $d \in [0,1]$ is a constant.

\subsection{Macroscopic changes}
\begin{figure}
    \begin{subfigure}[t]{0.48\columnwidth}
        \centering
        \includegraphics[width=\textwidth]{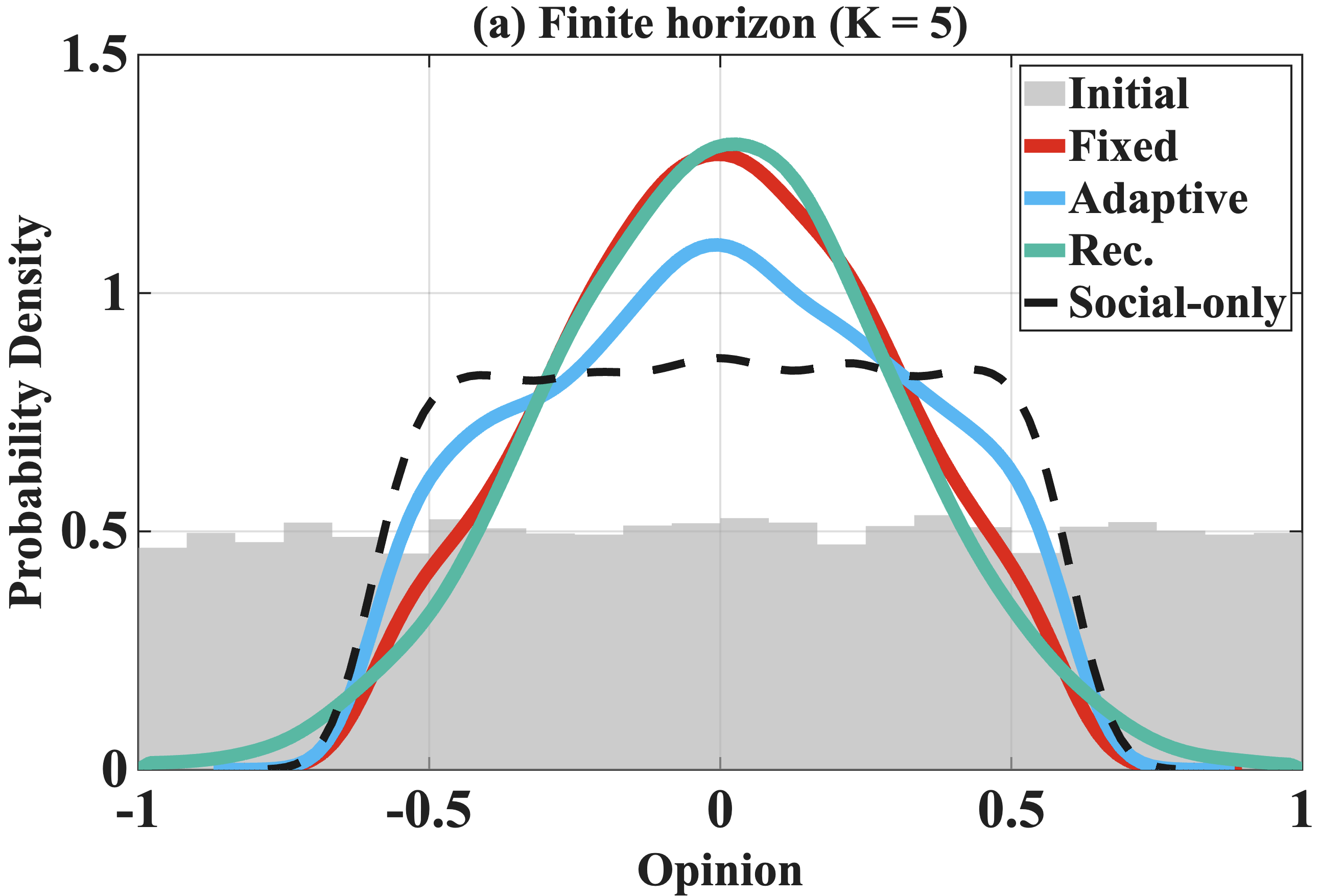}
        \caption{Finite horizon ($K=5$)}
    \end{subfigure}
    \hfill
    \begin{subfigure}[t]{0.48\columnwidth}
        \centering
        \includegraphics[width=\textwidth]{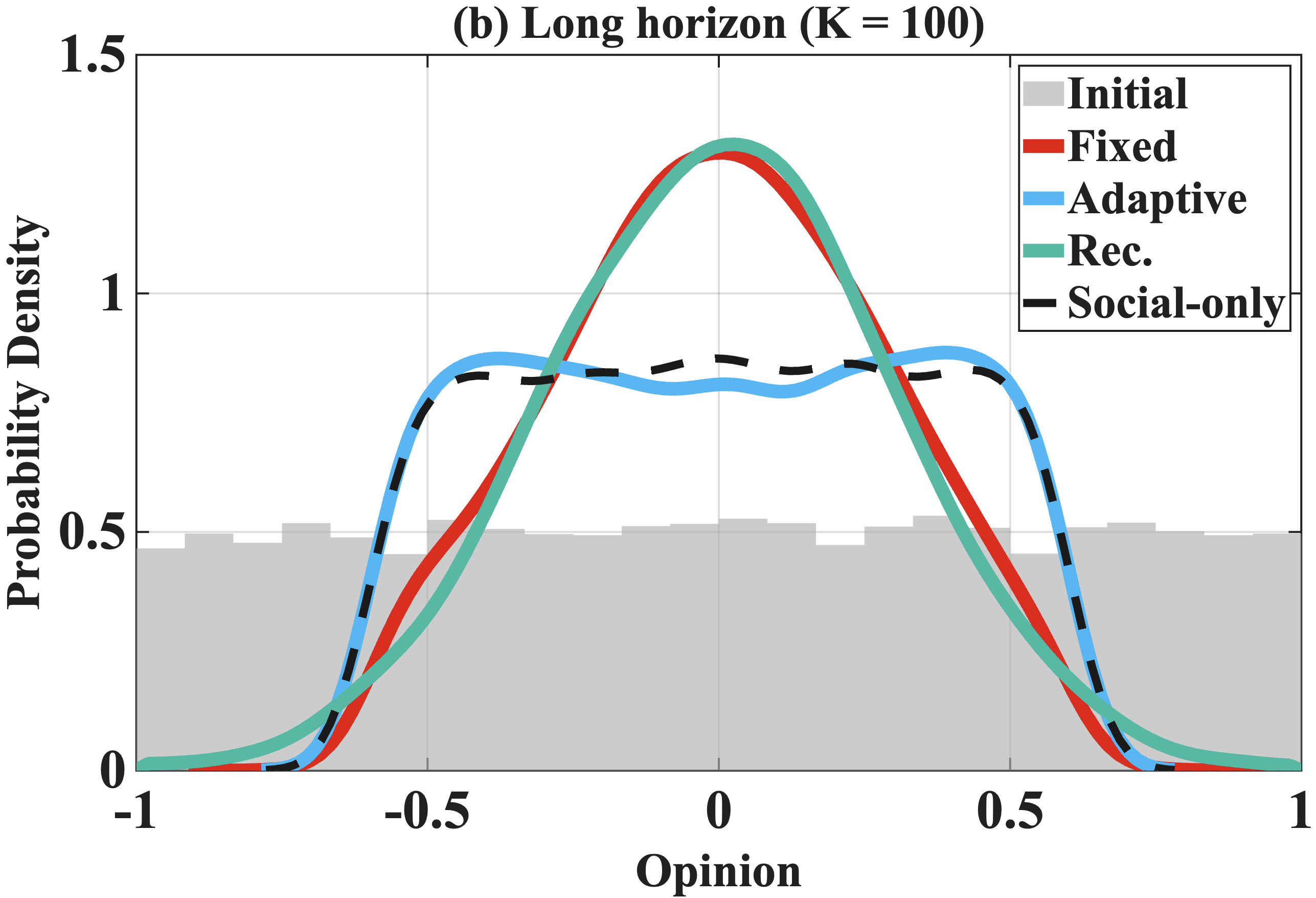}
        \caption{Long horizon ($K=100$)}
    \end{subfigure}
    \caption{Final opinion distribution under each agent policy after (a) a short, finite horizon and (b) a long horizon.}\label{fig:1-population}
    \vspace{-0.1in}
\end{figure}

Figure~\ref{fig:1-population} illustrates the impact of each policy on the distribution of the agents' final opinion when the platform follows the \emph{fixed recommendation} policy. Parameters are $\alpha = 0.3, \beta = 0.2, \gamma_0 = 0.6, \kappa = 1.2, \delta = 0.2$ and $K=5$ and $K=100$ for a short and long horizon, respectively. We consider a population of $10,000$ agents with innate opinions $x_0$ uniformly distributed over $[-1,1]$, interacting over a uniform social influence matrix
$W=\frac{1}{n}\mathbf 1\mathbf 1^\top$, with each agent $i$ receiving a fixed recommendation $u_{i,0}$ drawn from a zero-mean Gaussian distribution. The opinion distribution under the \emph{fixed} policy moves substantially toward the recommendation distribution, reflecting repeated direct exposure to recommendations and the propagation of these opinion shifts through the network via the social influence term $Wx_k$. Under the \emph{adaptive decreasing} policy, agents may reduce their clicking due to opinion drifts, which limits the \emph{direct} influence of the platform on their opinions. However, opinions do not generally revert to the initial distribution, because social influence persists even when agents do not click, and agents can still be affected \emph{indirectly} through neighbors whose opinions continue to evolve under recommendations. This is highlighted by the \emph{social-only} baseline (no click always): the adaptive policy’s distribution stays close to this social-only baseline, indicating that the remaining drift is primarily due to persistent social influence rather than continued direct exposure to recommendations. The mild concentration near the center at the short horizon occurs because for some agents the deviation threshold is rarely triggered (e.g., when their assigned $u_{i,0}$ is close to $x_{i,0}$), so the \emph{fixed} and \emph{adaptive decreasing} policies behave similarly.

\begin{figure*}[t]
    \centering
    \begin{subfigure}[t]{0.24\textwidth}
        \centering
        \includegraphics[width=1\textwidth]{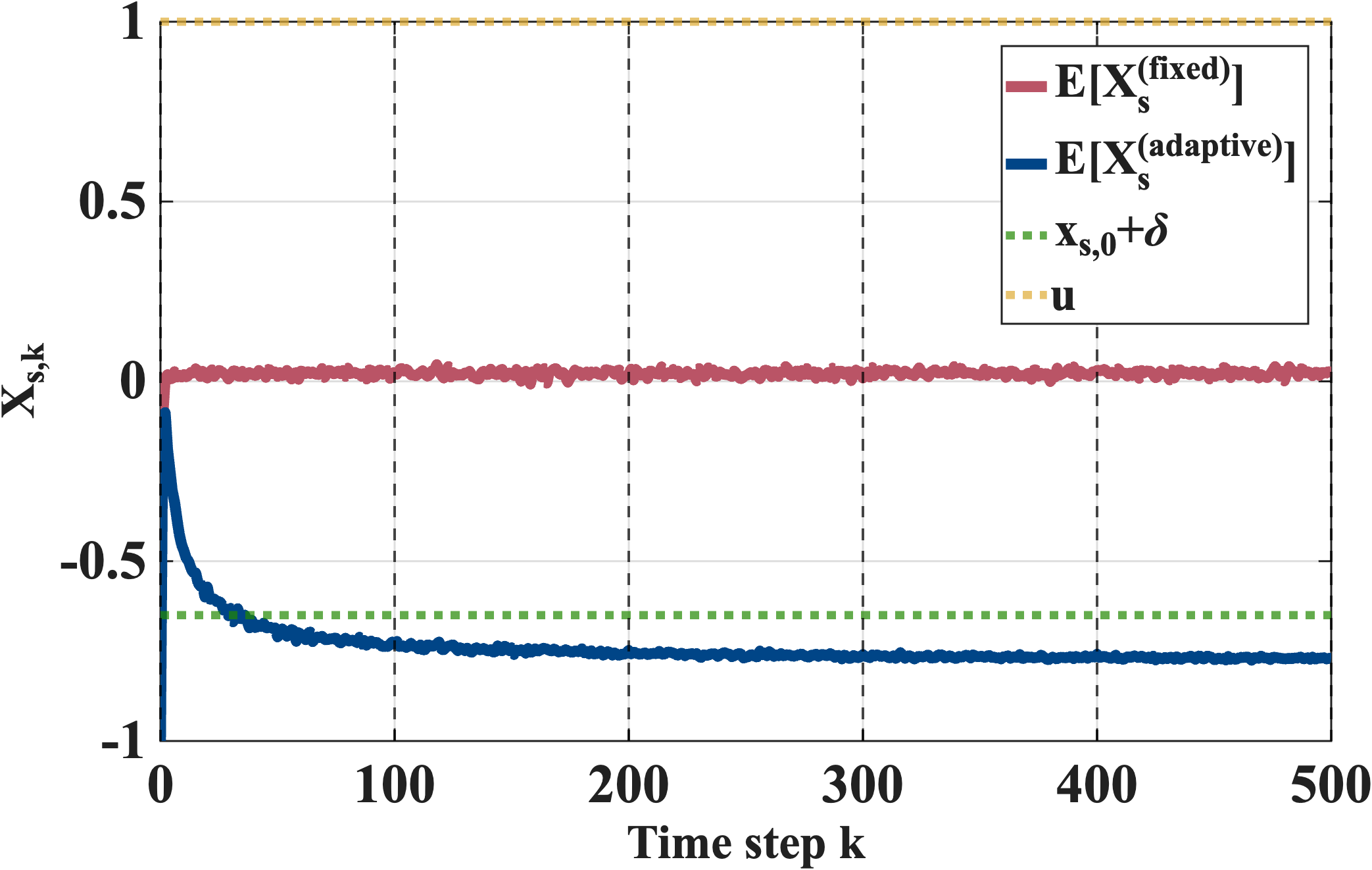}
        \caption{Strategic agent's expected opinion.}
        \label{fig:2-compare-rec-fixed-opinion}
    \end{subfigure}
    \hfill
    \begin{subfigure}[t]{0.24\textwidth}
        \centering
        \includegraphics[width=1\textwidth]{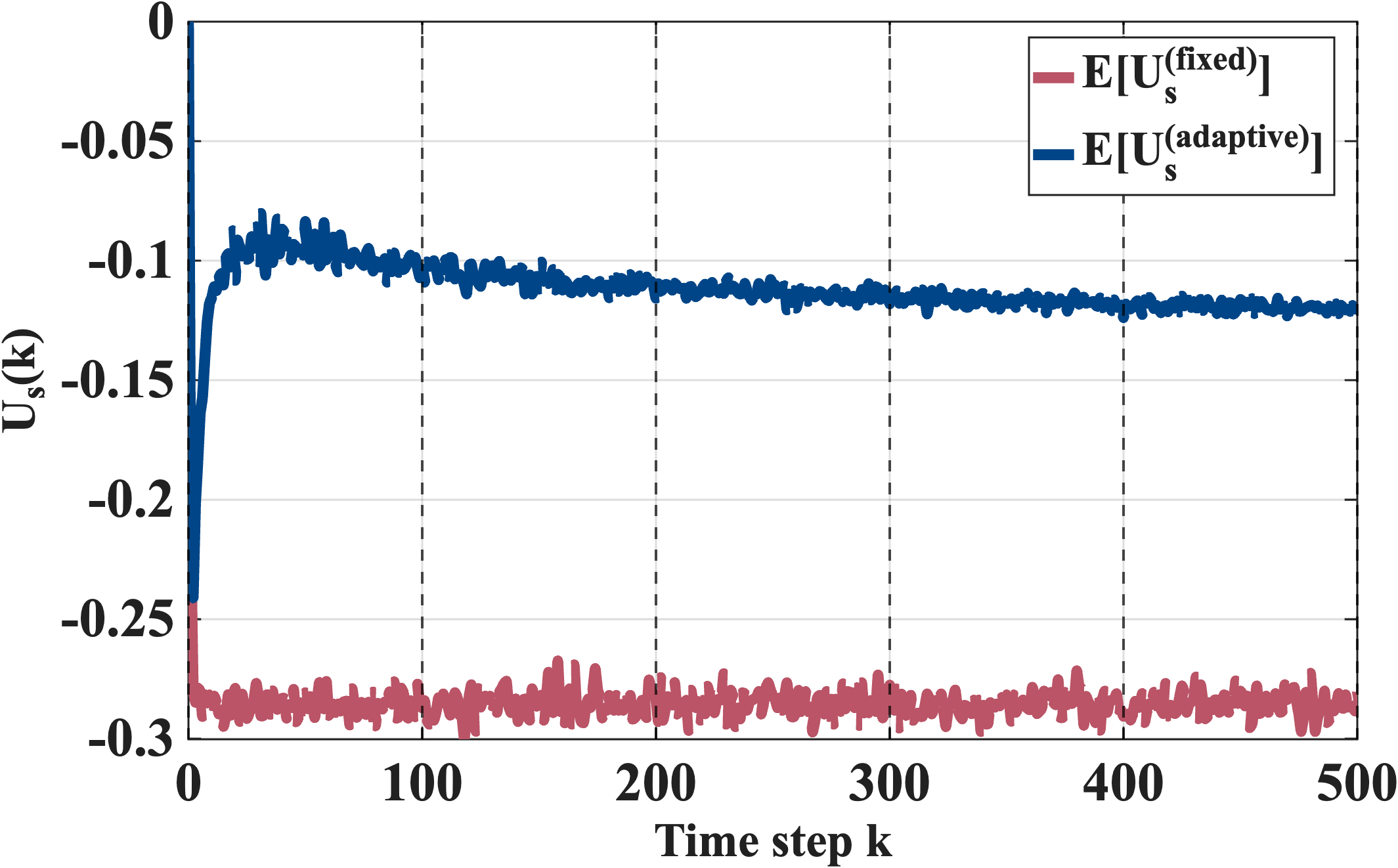}
        \caption{Strategic agent's expected utility.}
        \label{fig:2-compare-rec-fixed-agent}
    \end{subfigure}
    \hfill
    \begin{subfigure}[t]{0.24\textwidth}
        \centering
        \includegraphics[width=1\textwidth]{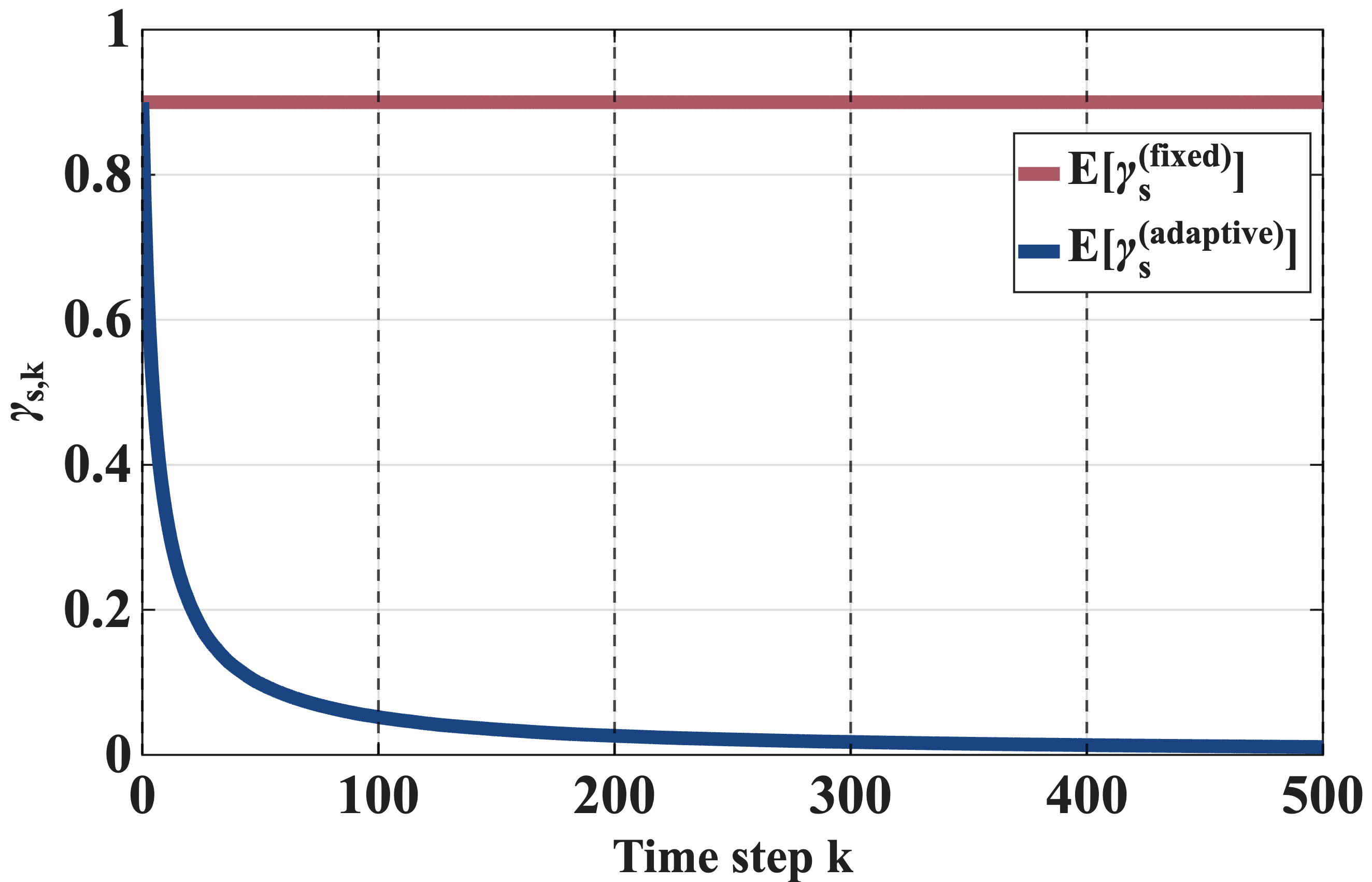}
        \caption{Strategic agent's expected clicking probability.}
        \label{fig:2-compare-rec-fixed-gamma}
    \end{subfigure}
    \hfill
    \begin{subfigure}[t]{0.24\textwidth}
        \centering
        \includegraphics[width=1\textwidth]{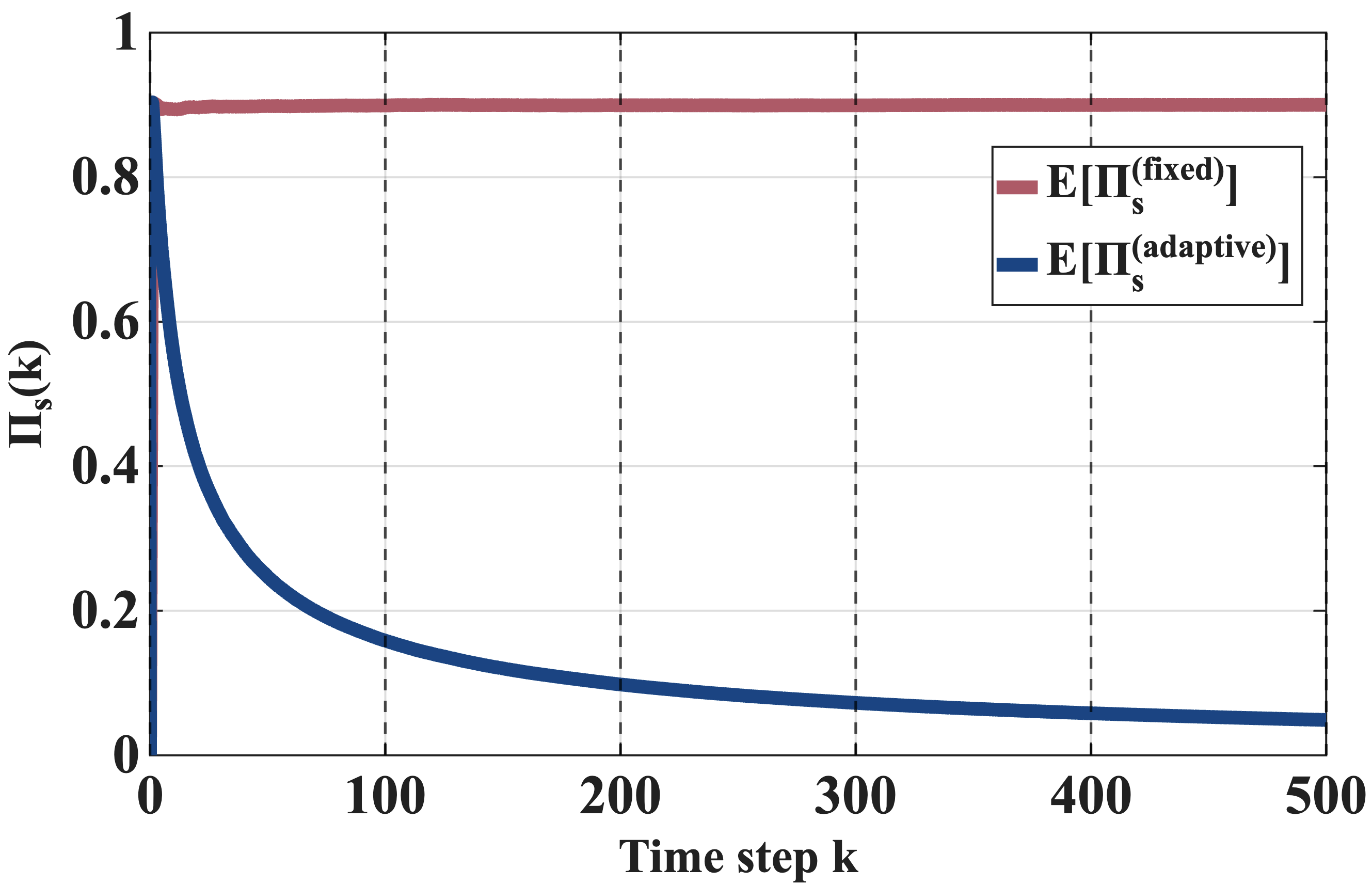}
        \caption{Platform's expected utility toward the strategic agent.}
        \label{fig:2-compare-rec-fixed-plat}
    \end{subfigure}

    \caption{Strategic agent's opinion, agent's utility, platform's utility, and agent's clicking probability under fixed recommendations, with social influence acting through population averaging and no outgoing influence from the strategic agent to others ($W_{-s,s}=0$).}
    \label{fig:compare-fixed-rec}
\end{figure*}

\begin{figure*}[th]
    \centering
    \begin{subfigure}[t]{0.24\textwidth}
        \centering
        \includegraphics[width=1\linewidth]{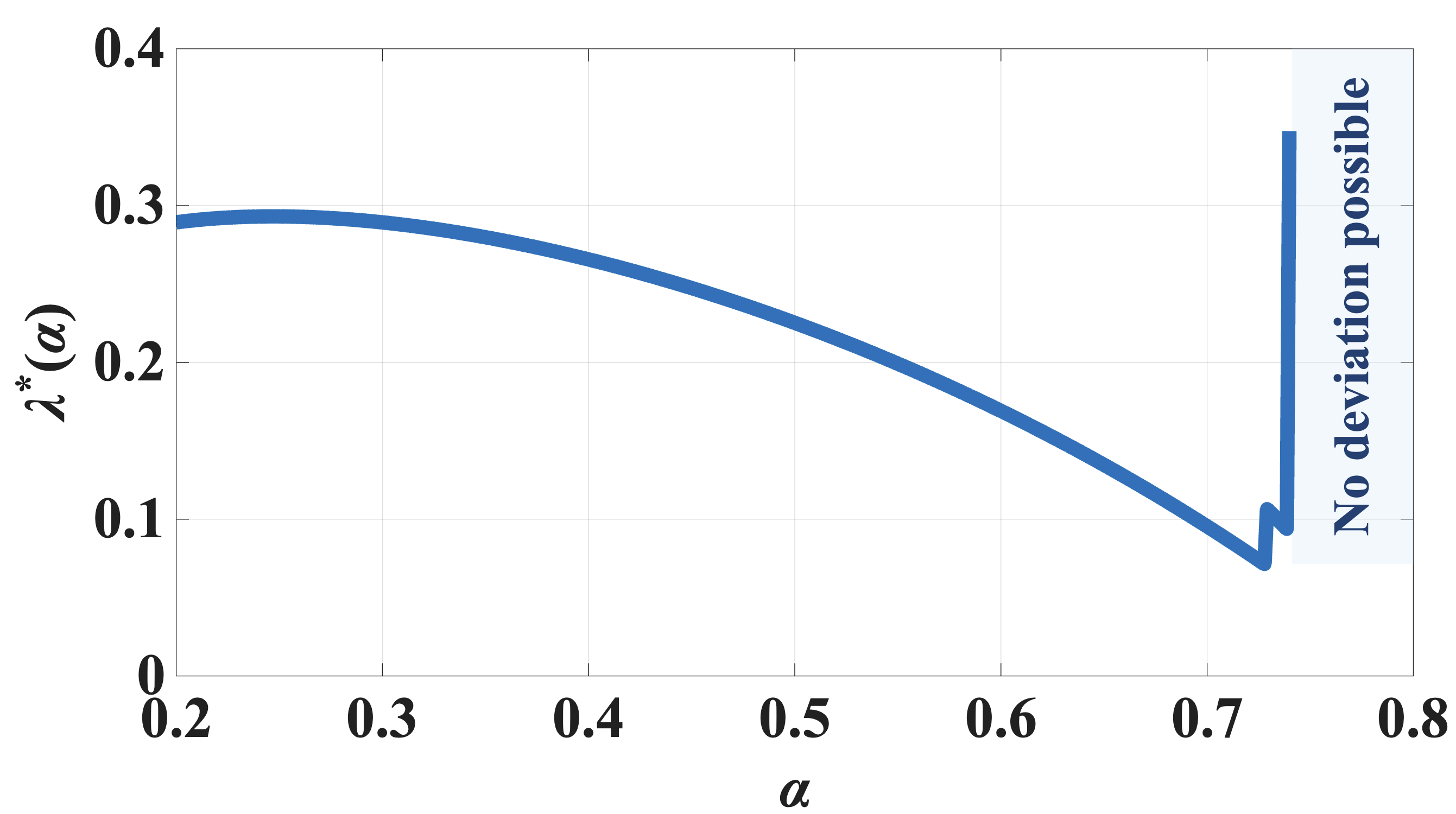}
        \caption{Guaranteed $\lambda$ range vs.\ $\alpha$}
        \label{fig:4-lambda-vary-alpha}
    \end{subfigure}
    \hfill
    \begin{subfigure}[t]{0.24\textwidth}
        \centering
        \includegraphics[width=1\linewidth]{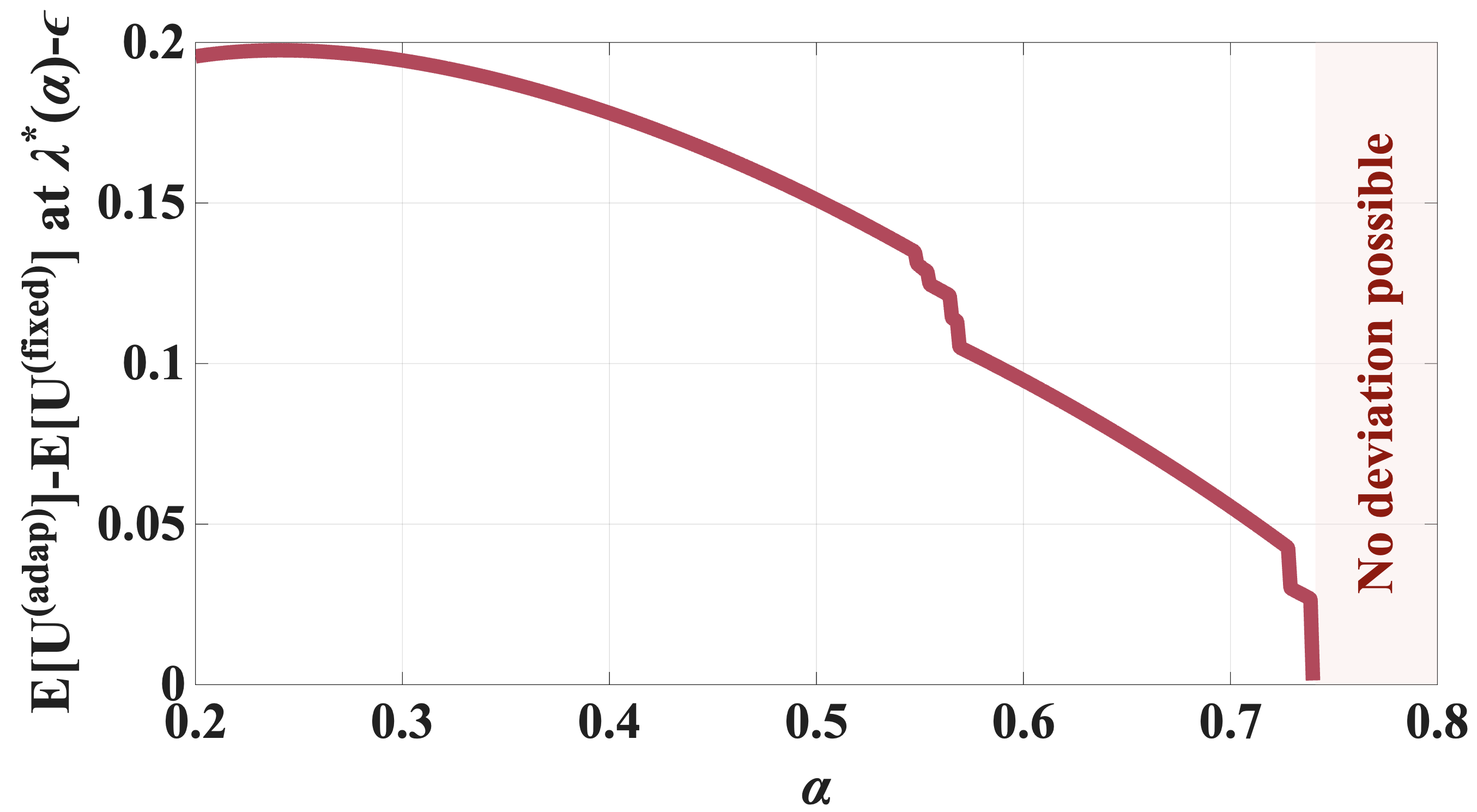}
        \caption{Utility difference near bound.}
        \label{fig:4-lambda-utility-diff-vary-alpha}
    \end{subfigure}
    \hfill
    \begin{subfigure}[t]{0.24\textwidth}
        \centering
        \includegraphics[width=1\linewidth]{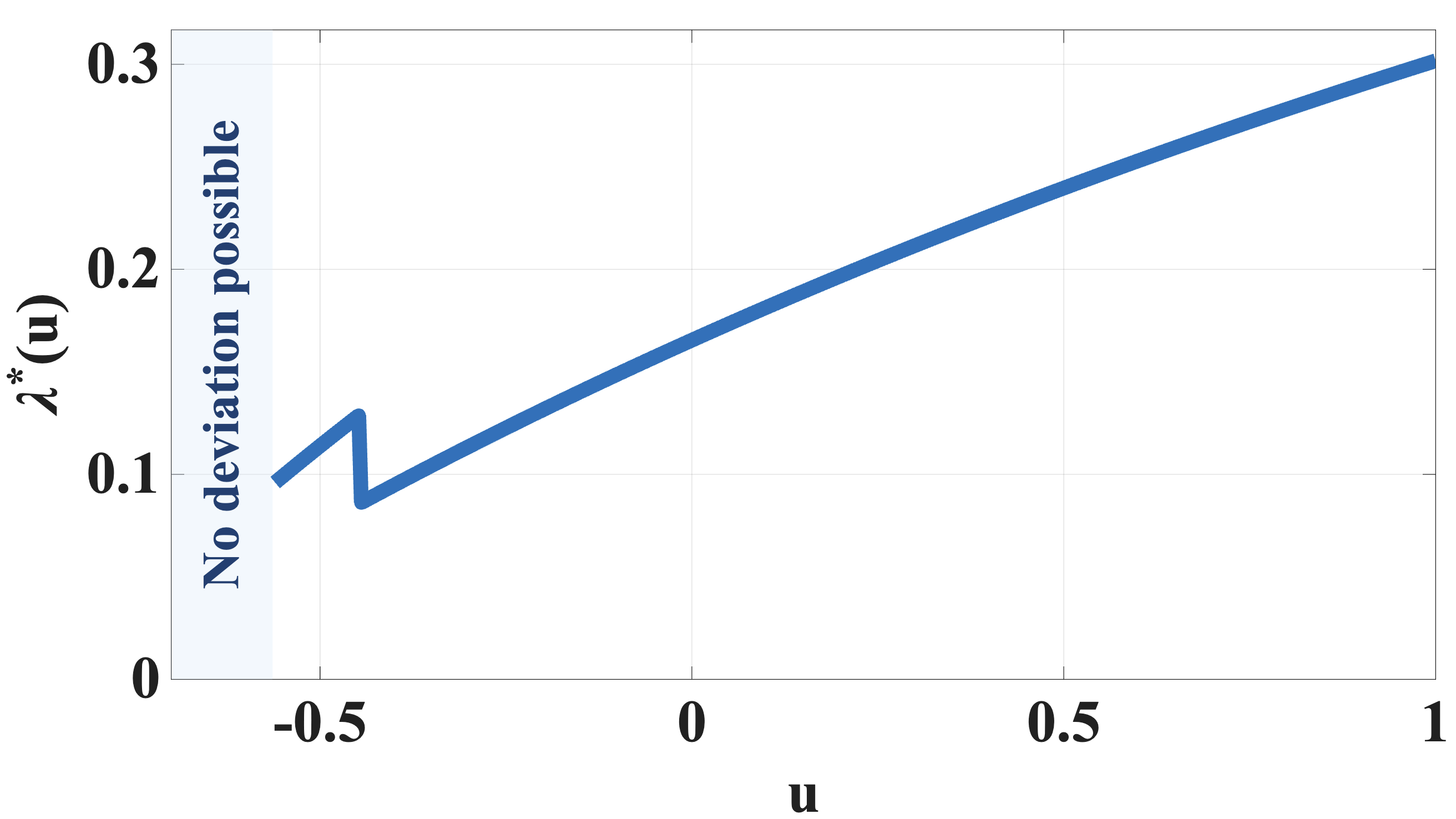}
        \caption{Guaranteed $\lambda$ range vs.\ $u$.}
        \label{fig:4-lambda-vary-u}
    \end{subfigure}
    \hfill
    \begin{subfigure}[t]{0.24\textwidth}
        \centering
        \includegraphics[width=1\linewidth]{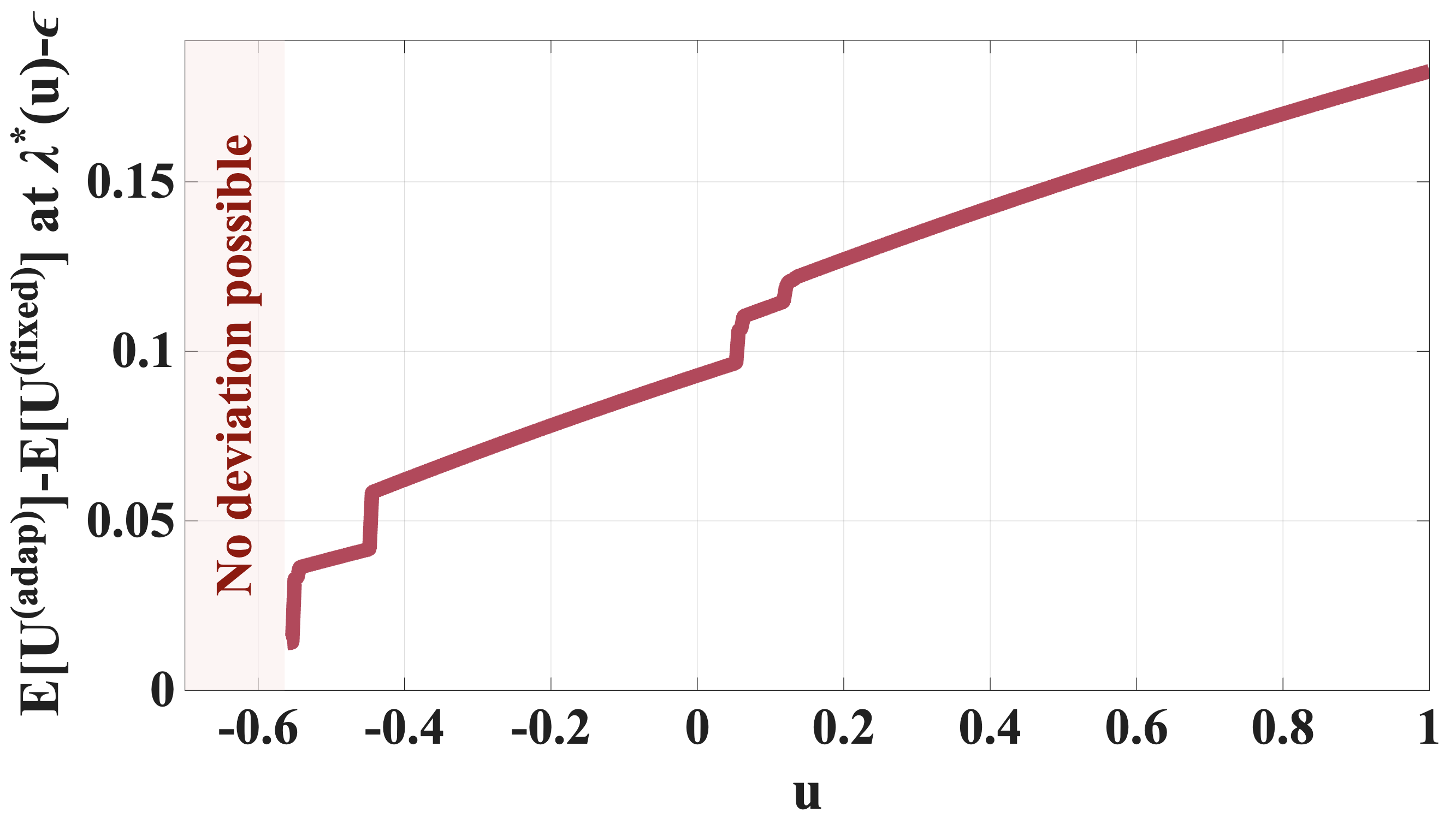}
        \caption{Utility difference near bound.}
        \label{fig:4-lambda-utility-diff-vary-u}
    \end{subfigure}
    \caption{The guaranteed range $\lambda < \lambda^*$ where the adaptive policy is theoretically proven (Proposition~\ref{prop:finite-adaptive-better-than-fixed}) to outperform the fixed policy in expected utility, (a) while varying $\alpha$ and (c) while varying $u$, verified numerically ((b) and (d)) near the bound (at $\lambda^*-\varepsilon$).
    }
    \label{fig:lambda-adap-outperforms}
\end{figure*}

\begin{figure*}[th]
    \centering
    \begin{subfigure}[t]{0.24\textwidth}
        \centering
        \includegraphics[width=1\textwidth]{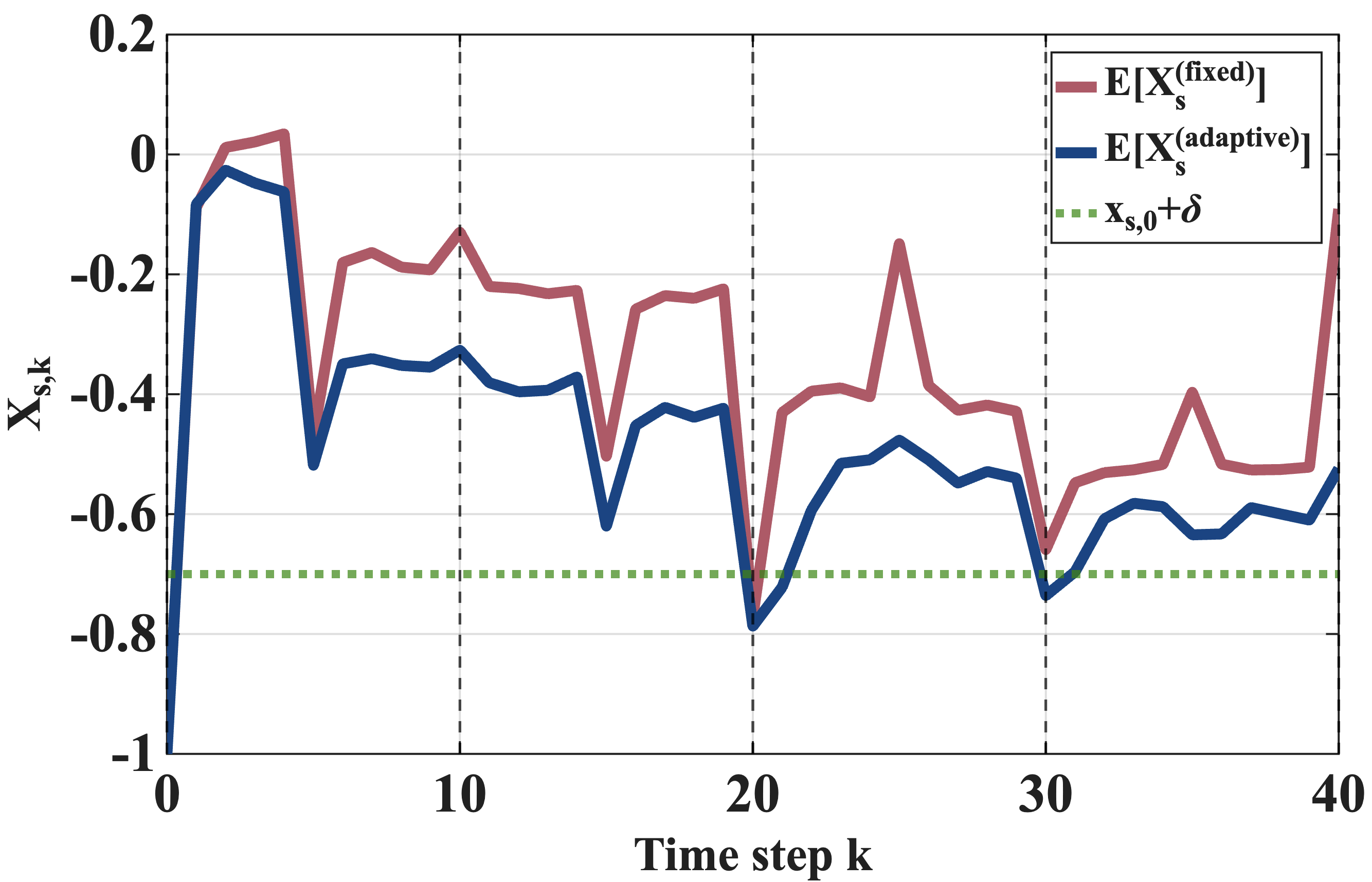}
        \caption{Agent's expected opinion.}
        \label{fig:3-compare-rec-eps-opinion}
    \end{subfigure}
    \hfill
    \begin{subfigure}[t]{0.24\textwidth}
        \centering
        \includegraphics[width=1\textwidth]{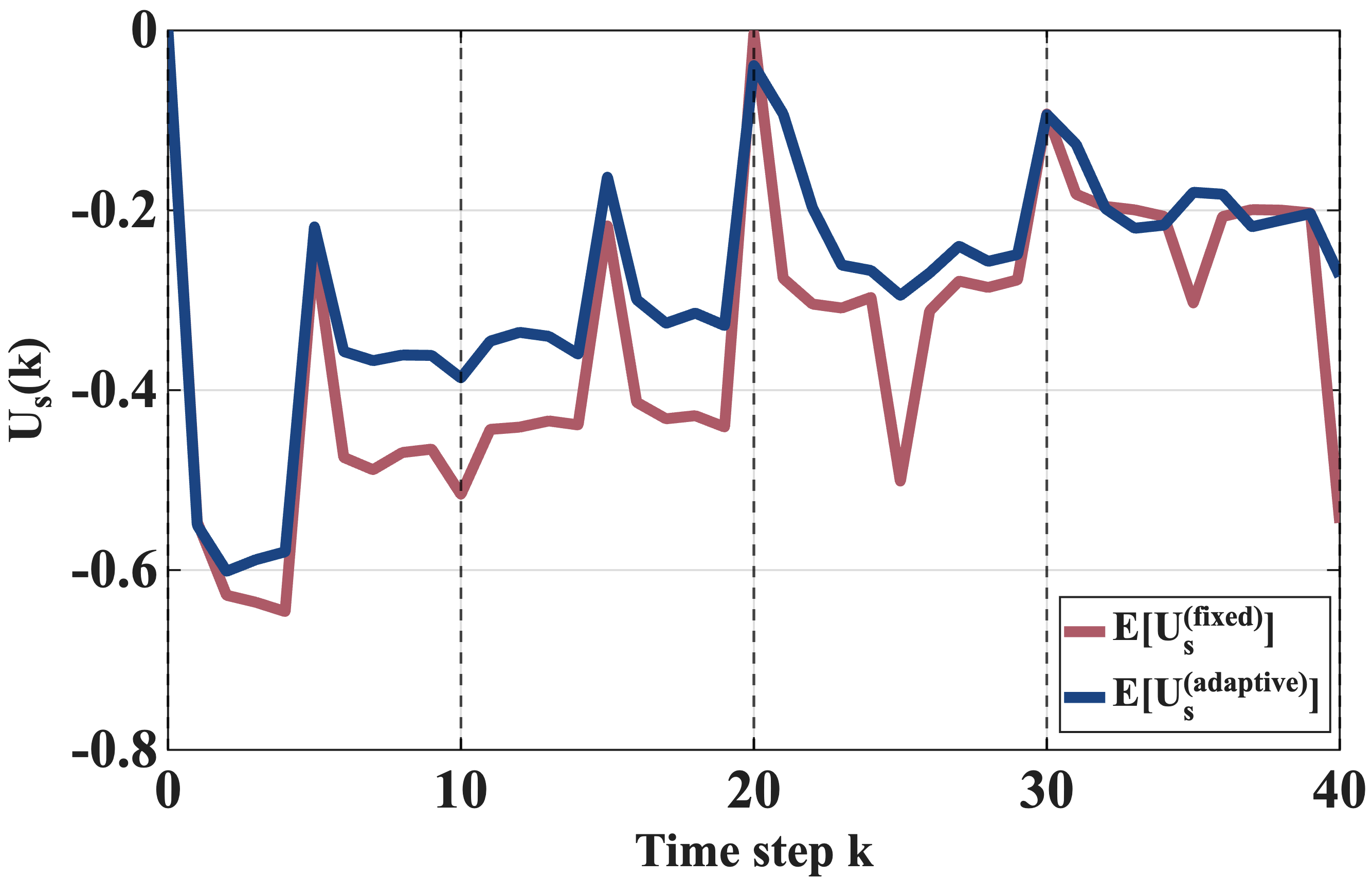}
        \caption{Agent's expected utility.}
        \label{fig:3-compare-rec-eps-agent}
    \end{subfigure}
    \hfill
    \begin{subfigure}[t]{0.24\textwidth}
        \centering
        \includegraphics[width=1\textwidth]{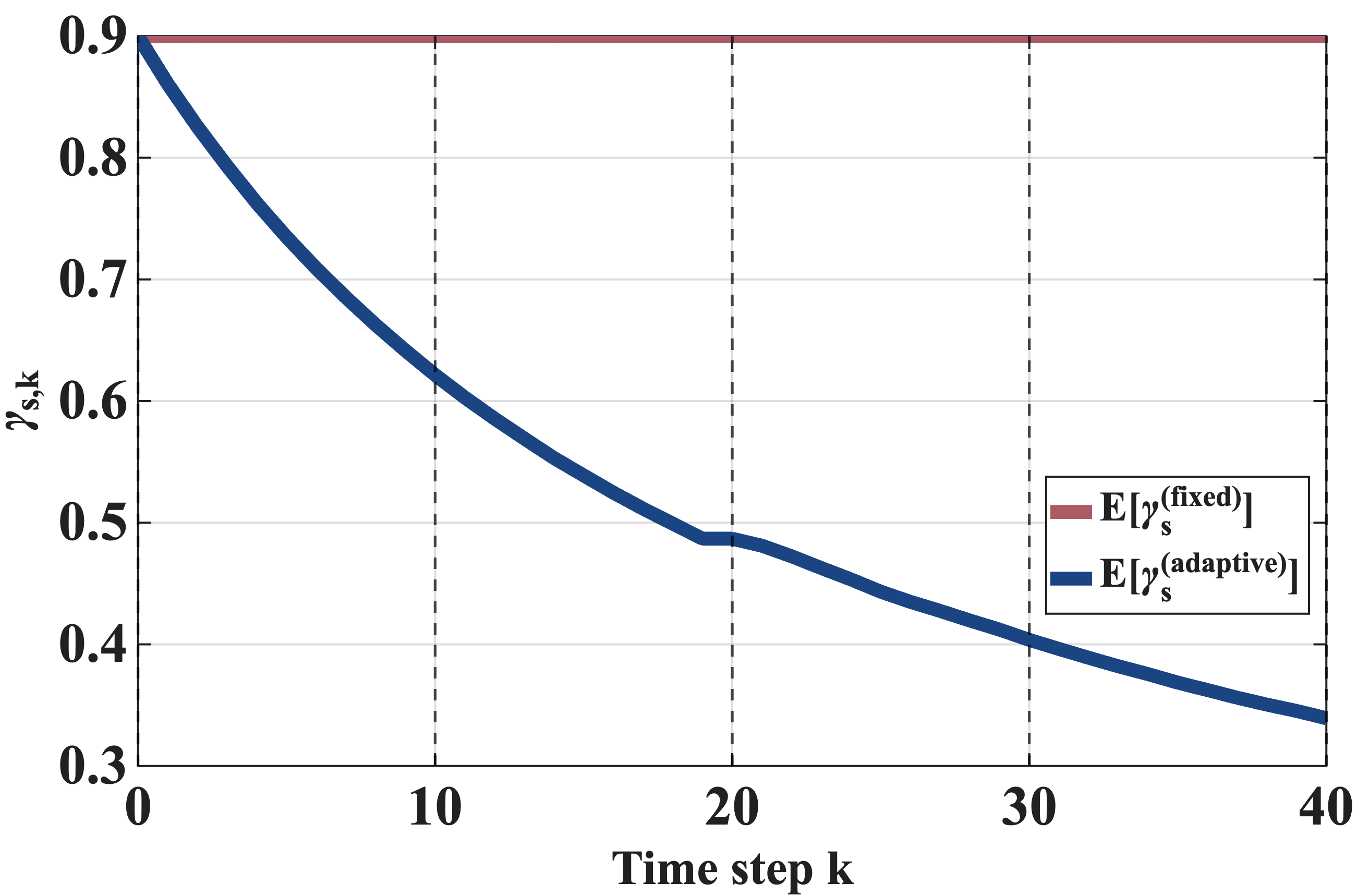}
        \caption{Expected clicking probability.}
        \label{fig:3-compare-rec-eps-gamma}
    \end{subfigure}
    \hfill
    \begin{subfigure}[t]{0.24\textwidth}
        \centering
        \includegraphics[width=1\textwidth]{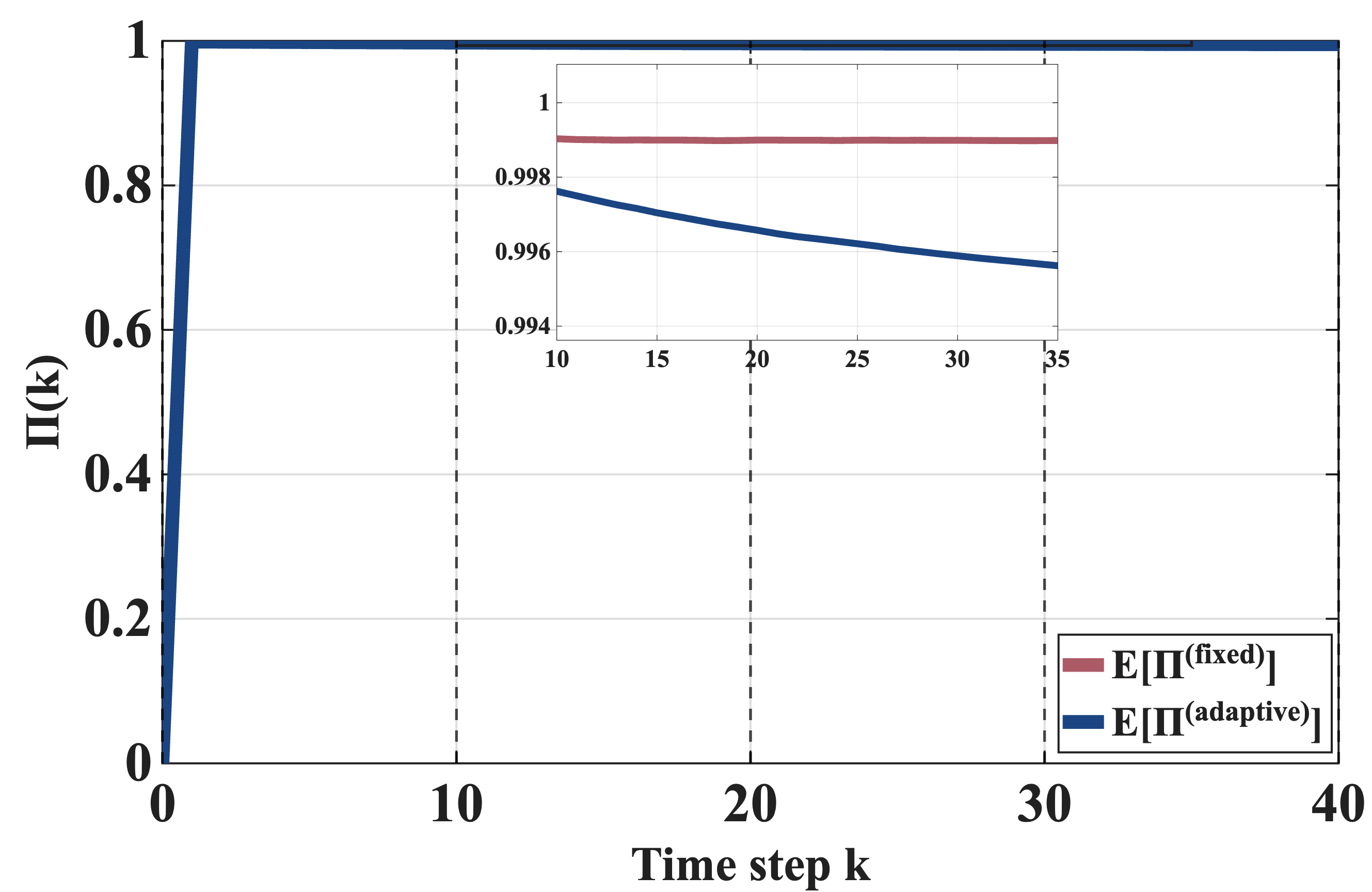}
        \caption{Platform's expected utility.}
        \label{fig:3-compare-rec-eps-plat}
    \end{subfigure}

    \caption{Strategic agent's opinion, utility, and clicking probability, and platform's utility, under varying recommendations.}
    \label{fig:compare-vary-rec}
\end{figure*}

\subsection{Comparing passive and reactive policies}
Figure~\ref{fig:compare-fixed-rec} shows the strategic agent's expected opinion (\ref{fig:2-compare-rec-fixed-opinion}), expected utility (\ref{fig:2-compare-rec-fixed-agent}), expected clicking probability (\ref{fig:2-compare-rec-fixed-gamma}), and the platform’s expected utility toward the strategic agent (\ref{fig:2-compare-rec-fixed-plat}), at each time step over a horizon of $K = 500$, when the platform follows the \emph{fixed recommendation}. We estimate each expectation by averaging over $1000$ Monte Carlo trials in a mixed population of $100$ agents: all agents $i\neq s$ click at every time step, while agent $s$ follows either a fixed clicking rate $\gamma_0$ or the adaptive decreasing policy. We use $\alpha=0.4$, $\beta=0.1$, $\lambda=0.4$, $\gamma_0=0.9$, $\kappa=1.2$, $\delta=0.35$, and $d=0.1$. Initial opinions and recommendations are fixed at $x_{i,0}=-1$ and $u_{i,0}=1$ for all agents.

From Fig.~(\ref{fig:2-compare-rec-fixed-opinion}), we observe that under the \emph{fixed} clicking policy, its expected opinion gradually drifts toward the recommendation, this shift is further propagated through the network, and stabilizes at a higher long-term value (Corollary~\ref{cor:expectation-opinion-mixed-fixed-infinity}). In contrast, under the \emph{adaptive decreasing} clicking policy, each time the deviation threshold is crossed by the strategic agent, the clicking probability is reduced, leading to long-term disengagement and reversion of the opinion $x_{s,k}$ closer to the no click trajectory rather than tracking the recommendation (Corollary~\ref{cor:expectation-opinion-adaptive-infinity}).

From Fig.~(\ref{fig:2-compare-rec-fixed-agent}), we observe that the strategic agent reaches higher (lower) expected utility under the \emph{adaptive decreasing} (\emph{fixed}) policy, as the former better preserves the innate opinion and places more weight on limiting opinion drift rather than maximizing content consumption. On the other hand, the platform's expected utility from agent $s$, shown in Fig.~(\ref{fig:2-compare-rec-fixed-plat}), follows the clicking behavior, being higher under the \emph{fixed} policy and lower under the \emph{adaptive decreasing} policy as $\gamma_{s,k}$ decreases. Finally, Fig.~(\ref{fig:2-compare-rec-fixed-gamma}) supports the findings of Proposition~\ref{prop:gamma-convergence} by confirming the convergence of clicking probability of the \emph{adaptive policy} to zero in the long run.

\subsection{The adaptive policy can outperform the fixed policy in finite time horizons} We next verify the findings of Proposition~\ref{prop:finite-adaptive-better-than-fixed}. In these experiments, we consider a short horizon of $K = 5$, fix $\beta = 0.2, \kappa = 1.2, \delta = 0.3$, and fix a network $W$ and an initial opinion profile $x_0$ that satisfy Assumptions~\ref{ass:no-outgoing-s} and ~\ref{ass:sign-preservation}. Throughout, all agents $i\neq s$ click at every time step, while agent $s$ follows either the fixed policy~\ref{alg:policy-one} or the adaptive decreasing policy~\ref{alg:policy-two} with $\gamma_{s,0} = 1$, in either cases. Recall that Proposition~\ref{prop:finite-adaptive-better-than-fixed} provides a sufficient threshold on the value of $\lambda$ (the preference for content consumption in the agent's utility) under which the \emph{adaptive decreasing} policy is guaranteed to outperform the \emph{fixed} policy in expected utility, in a finite horizon. Figures~\ref{fig:4-lambda-vary-alpha} and~\ref{fig:4-lambda-vary-u} plot these sufficient bounds, $\lambda^*(\alpha)$ and $\lambda^*(u)$, as a function of $\alpha$ and $u$, respectively. We next numerically evaluate the agent's expected utilities under each policy near these bounds, by setting at $\lambda=\lambda^*(\cdot)-\varepsilon$, where $\varepsilon=0.02$.

In one experiment, we fix $u = 1$ and vary $\alpha \in [\beta, 1-\beta]$ (Fig.~\ref{fig:4-lambda-utility-diff-vary-alpha}), and in a second experiment, we fix $\alpha = 0.3$ and vary the recommended content $u \in [-0.7,1]$ (Fig.~\ref{fig:4-lambda-utility-diff-vary-u}). The plots show the utility difference between the adaptive and fixed policies, averaged over $5000$ Monte Carlo trials, verifying that the expected utility under the adaptive policy is higher than that of the fixed policy in the short term for the values of $\lambda$ just below the theoretical threshold in Proposition~\ref{prop:finite-adaptive-better-than-fixed}. Note that in both cases, the shaded region indicates parameter values for which no deviation is triggered within the horizon (formally, $M$ exceeds $K$). In that scenario, the adaptive policy never reduces its clicking probability and therefore behaves identically to the fixed policy and the utility gap approaches zero.

\subsection{Exploring recommendations}
We next consider a platform following the \emph{explore periodically} policy described in Section~\ref{sec:platform-policy}. Figure~\ref{fig:compare-vary-rec} shows the expected opinion of the strategic agent $s$ (\ref{fig:3-compare-rec-eps-opinion}), its expected utility (\ref{fig:3-compare-rec-eps-agent}), its expected clicking probability (\ref{fig:3-compare-rec-eps-gamma}), and the platform's expected utility measured by the population CTR (\ref{fig:3-compare-rec-eps-plat}), over a horizon of $K = 40$. We use an exploration period of $T=5$ (the platform samples a new $u_k$ uniformly at random every 5 steps). Results are averaged over $1000$ independent Monte Carlo trials, in a population of $99$ passive agents (who click at every time step) and one strategic agent $s$. We let $x_{i,0} =-1, \forall i$, $\alpha = 0.4, \beta = 0.1, \lambda = 0.2, \gamma_0 = 0.9, \kappa = 1.05, \delta = 0.3, n = 100$ and $d=0$. 

Under the \emph{explore periodically} policy, exploration steps introduce occasional new changes in the recommendations, which can temporarily pull the reactive agent's opinion away from its innate value. If an explored recommendation achieves the highest empirical CTR so far, it is subsequently exploited, which can shift the population's opinion toward that recommendation. If the explored recommendation is not aligned with the reactive agents' opinion (i.e., pulls its opinion further away from the innate opinion), then the \emph{fixed} policy~\ref{alg:policy-one} performs as before since the strategic agent keeps clicking with the fixed probability of $\gamma_0$ (and all other agents always click). In contrast, under the \emph{adaptive decreasing} policy~\ref{alg:policy-two}, there may be reductions in clicking probability at those steps, keeping the expected opinion closer to $x_{s,0}$ than under the \emph{fixed} policy~\ref{alg:policy-one} (Fig.~\ref{fig:3-compare-rec-eps-opinion}). When the platform explores, it is more probable to find a new recommendation that matches the agents' preferences (relative to the platform's \emph{fixed recommendation} policy); in that case the deviation threshold $\delta$ is triggered less often, so the differences between the two strategic agent's policies are smaller, leading to closer agent's expected utilities than in the \emph{fixed recommendation} case (Fig.~\ref{fig:3-compare-rec-eps-agent}).
The platform's expected utility under the two policies is very close; the zoomed view highlights the small but consistent separation that is hard to see at the full scale. The is because with $100$ agents and $99$ of them always clicking, the population CTR is only weakly affected by the behavior of a single strategic agent (Fig.~\ref{fig:3-compare-rec-eps-plat}). Figure~\ref{fig:3-compare-rec-eps-gamma} also shows that $\gamma_{s,k}$ stays constant under the fixed policy, while it decreases over time only under the adaptive policy. Finally, unlike the platform's \emph{fixed recommendation} policy, the \emph{explore periodically} policy reductions do not necessarily occur as frequently, as the platform can hone in on recommendations closer to the agent's opinion by leveraging exploration.
\section{Conclusion}\label{sec:conclusions}

We studied the opinion dynamics of \emph{reactive} agents in a social recommender system, where an agent can decide whether to consume the recommended content based on its evolving opinions, with the goal of limiting opinion drifts. Through analytical and numerical experiments, we demonstrated that reactive agents can limit persuasion toward a platform's recommendation by adaptively adjusting their content consumption behavior, even when this may still be limited by the indirect, non-vanishing effect of social influences. These findings help to better understand how user-level strategies can challenge biases induced by recommendation algorithms. Main directions of future research include analytical studies of macroscopic effects (opinion distributions), closed-loop platform policies that adapt recommendations in response to opinion shifts, and alternative stochastic agent policies, such as not only decreasing but also increasing the clicking probability.

\bibliographystyle{IEEEtran}
\bibliography{references}

\appendix

\makeatletter
\def\@seccntformat#1{%
  \ifcsname the#1\endcsname
    \csname the#1\endcsname\quad
  \fi}
\makeatother

\setcounter{subsection}{0}
\renewcommand{\thesubsection}{\arabic{subsection}.}

\subsection{Proof of Lemma~\ref{lem:convex-combination-opinion}}

\begin{proof}
    We proceed by substituting $x_k$ recursively into \eqref{eq:user-model-PQ}, which yields
    \begin{align*}
        x_k &= (I - Q_{k-1} - P_{k-1}) x_0 + Q_{k-1}W x_{k-1} + P_{k-1}u_0\\
        &= (I - Q_{k-1} - P_{k-1}) x_0 + P_{k-1}u_0\\
          &\quad + Q_{k-1}W\big((I - Q_{k-2} - P_{k-2})x_0 + P_{k-2}u_0 + Q_{k-2}W x_{k-2}\big)\\
        &= \big((I - Q_{k-1} - P_{k-1}) + Q_{k-1}W (I - Q_{k-2} - P_{k-2})\big)x_0\\
          &\quad + \big(P_{k-1} + Q_{k-1}W P_{k-2}\big)u_0
          + (Q_{k-1}W)(Q_{k-2}W)x_{k-2},
    \end{align*}
    and continuing in this manner gives the following pattern:
    \[x_k \;=\; M_k x_0 \;+\; G_k u_0\]
    where
    \begin{align*}
        M_k \;&=\; \Phi(k,0)+\sum_{t=0}^{k-1} \Phi(k,t+1)\,(I-Q_t-P_t),\\
        G_k \;&=\;\sum_{t=0}^{k-1} \Phi(k,t+1)\,P_t.
    \end{align*}
     and for $t\in\{0,\dots,k-1\}$,
    \[
    \Phi(k,t+1) := \prod_{\tau=t+1}^{k-1} (Q_{\tau} W),
    \qquad
    \Phi(k,k):=I,
    \]

    Now, we show by induction that $S_k:=M_k\mathbf{1} + G_k\mathbf{1}=\mathbf{1}$ for all $k$.
    We can write the one-step recursion in $(M_k,G_k)$ form:
    \begin{align*}
        x_{k+1}
        &= (I-Q_k-P_k) x_0 + P_k u_0 + Q_k W (M_k x_0 + G_k u_0)\\
        &= ((I-Q_k-P_k) + Q_k W M_k)x_0 + (P_k + Q_k W G_k)u_0.
    \end{align*}
    Hence
    \[
    M_{k+1}=I-Q_k-P_k+Q_k W M_k,
    \qquad
    G_{k+1}=P_k+Q_k W G_k.
    \]
    Multiplying by $\mathbf{1}$ and adding gives
    \begin{align*}
        S_{k+1}
        &= (I-Q_k-P_k+Q_kWM_k)\mathbf{1} + (P_k+Q_kWG_k)\mathbf{1}\\
        &= (I-Q_k-P_k+P_k)\mathbf{1} + Q_kW(M_k\mathbf{1}+G_k\mathbf{1})\\
        &= (I-Q_k)\mathbf{1} + Q_kW S_k.
    \end{align*}
    Since $W$ is row-stochastic, $W\mathbf{1}=\mathbf{1}$, and thus if $S_k=\mathbf{1}$ then
    \[
    S_{k+1}=(I-Q_k)\mathbf{1}+Q_k\mathbf{1}=\mathbf{1}.
    \]
    For the base case, at $k=0$ we have $M_0=I$ and $G_0=0$, hence $S_0=\mathbf{1}$.
    Therefore $S_k=\mathbf{1}$ for all $k\geq0$, i.e.,
    \[
    M_k\mathbf{1}+G_k\mathbf{1}=\mathbf{1}.
    \]
    We have shown that for any realization of clicks,
    \[
    x_k = M_k x_0 + G_k u_0,
    \qquad\text{and}\qquad
    (M_k+G_k)\mathbf{1}=\mathbf{1}.
    \]
    Moreover, $W$ is nonnegative and row-stochastic, and $Q_t,P_t$ are non-negative diagonal matrices; hence each product $\Phi(k,t+1)$ has non-negative elements, which implies $M_k\ge 0$ and $G_k\ge 0$ element-wise. Therefore, for each $i$, the $i$-th row of $M_k$ and $G_k$ has non-negative weights that sum to $1$, so $x_{i,k}$ is a convex combination of the entries of $x_0$ and $u_0$. In particular, since $x_0,u_0\in[-1,1]^n$, it follows that $x_k\in[-1,1]^n$.
    \end{proof}

\subsection{Proof of Proposition~\ref{prop:expectation-opinion-fixed}}\label{app:expectation-opinion-fixed}

We begin by proving one lemma that will be used in the proof of Proposition~\ref{prop:expectation-opinion-fixed}.
\setcounter{lemma}{2}
\begin{lemma}\label{lem:invertible-geometric}
    Let $W$ be row-stochastic and let $q\in[0,1)$. Then $I-qW$ is invertible. Moreover,
    \[
    \sum_{t=0}^{k-1}(qW)^t = (I-qW)^{-1}\big(I-(qW)^k\big), \qquad k\ge 1.
    \]
\end{lemma}

\begin{proof}
    Since $W$ is row-stochastic and nonnegative, $\|W\|_\infty=\max_i\sum_{j=1}^n |w_{ij}|=\max_i\sum_{j=1}^n w_{ij}=1$, hence $\|qW\|_\infty=q\|W\|_\infty=q<1$. To show that $I-qW$ is invertible, it suffices to show its null space is trivial. Let $v$ satisfy $(I-qW)v=0$. Then $v=qWv$, and therefore
    \[
    \|v\|_\infty = \|qWv\|_\infty \le \|qW\|_\infty \|v\|_\infty = q\|v\|_\infty.
    \]
    If $v\neq 0$, then $\|v\|_\infty>0$ and dividing both sides gives $1\le q$, contradicting $q<1$.
    Hence $v=0$, so $\mathrm{Null}(I-qW)=\{0\}$ and thus $I-qW$ is invertible.
    For the finite geometric-series identity, let $W':=qW$. Then
    \[
    (I-W')\sum_{t=0}^{k-1}(W')^t
    =\sum_{t=0}^{k-1}((W')^t-(W')^{t+1})
    = I-(W')^k.
    \]
    Multiplying by $(I-W')^{-1}$ yields
    \[
    \sum_{t=0}^{k-1}(qW)^t = (I-qW)^{-1}\big(I-(qW)^k\big).
    \]
\end{proof}

Now, we turn to the proof of Proposition~\ref{prop:expectation-opinion-fixed}.
\begin{proof}
    Under the fixed clicking policy, each agent clicks independently with constant probability $\gamma_0$,
    so $c_{i,k}\sim \mathrm{Bernoulli}(\gamma_0)$ i.i.d. over $i$ and $k$, and hence
    \[
    \mathbb{E}[C_k]=\gamma_0 I.
    \]
    Taking expectation of \eqref{eq:user-model-PQ} and using linearity of expectation gives
    \[\mathbb{E}[X_{k+1}^\text{all}] = \bigl(I-\mathbb{E}[Q_k]-\mathbb{E}[P_k]\bigr)x_0
        + \mathbb{E}[P_k]u_0
        + \mathbb{E}[Q_k W X_k^\text{all}].\]
    We next simplify each term. Using $\mathbb{E}[C_k]=\gamma_0 I$,
    \begin{align*}
        \mathbb{E}[Q_k]&=bI+(\beta-b)\mathbb{E}[C_k]
        =\bigl(b+(\beta-b)\gamma_0\bigr)I = qI,\\
        \mathbb{E}[P_k]&=(1-\zeta)\mathbb{E}[C_k]=(1-\zeta)\gamma_0 I = pI.
    \end{align*}
    Moreover, $Q_k$ depends only on $C_k$ (time $k$ clicks), while $X_k$ depends on $\{C_0,\dots,C_{k-1}\}$,
    so $Q_k$ is independent of $X_k$. 
    
    Therefore
    \[
    \mathbb{E}[Q_k W X_k^\text{all}] = \mathbb{E}[Q_k]\,W\,\mathbb{E}[X_k^\text{all}] = q\,W\,\mathbb{E}[X_k^\text{all}].
    \]
    Substituting into initial expectation equation yields the recursion:
    \[\mathbb{E}[X_{k+1}^\text{all}]
    \;=\; (1-q-p)x_0 \;+\; p u_0 \;+\; qW\,\mathbb{E}[X_k^\text{all}],
    \qquad \mathbb{E}[x_0]=x_0.\]
    Let $s:=(1-q-p)x_0+pu_0$. Then,
    \[
    \mathbb{E}[X_{k+1}^\text{all}] = s + q\;W\; \mathbb{E}[X_{k}^\text{all}].
    \]
    Iterating gives
    \[
    \mathbb{E}[X_{k}^\text{all}] = (qW)^k x_0 + \sum_{t=0}^{k-1}(qW)^t\,s.
    \]
    Note that $q=(1-\gamma_0)b+\gamma_0\beta\le \max\{b,\beta\}$. Under $\alpha\ge\beta$ and $0<\alpha+\beta\le 1$, we have $b=\frac{\beta}{\alpha+\beta}\le \frac12$ and $\beta\le \frac12$,
    and therefore $q\le \frac12<1$.
    
    By Lemma~\ref{lem:invertible-geometric}, $I-qW$ is invertible and $A_k:=\sum_{t=0}^{k-1}(qW)^t=(I-qW)^{-1}(I-(qW)^k)$, and hence
    \[
    \mathbb{E}[X_k^\text{all}] = (qW)^k x_0 + A_k\bigl((1-q-p)x_0+pu_0\bigr).
    \]
    Rearranging the $x_0$ and $u_0$ terms gives the equivalent form stated in the proposition:
    \[
    \mathbb{E}[X_k^\text{all}]
    = \big((qW)^k + (1-q)A_k\big)x_0 + pA_k\,(u_0-x_0).\]
\end{proof}

\subsection{Proof of Corollary~\ref{cor:expectation-opinion-fixed-no-social}}\label{app:expectation-opinion-fixed-no-social}

\begin{proof}
    Substituting $W=I$ in Proposition~\ref{prop:expectation-opinion-fixed}, we have:
    \begin{align*}
        \mathbb{E}[X_k^\text{all}] &=\Big(q^k I+(1-q)\frac{1-q^k}{1-q}I\Big)x_0
        +p\frac{1-q^k}{1-q}(u_0-x_0)\\
        &= x_0 + p\frac{1-q^k}{1-q}(u_0-x_0).
    \end{align*}
\end{proof}

\subsection{Proof of Corollary~\ref{cor:expectation-opinion-fixed-infinity}}\label{app:expectation-opinion-fixed-infinity}

\begin{proof}
    Consider the expression for $\mathbb{E}[X_k^\text{all}]$ found in Proposition~\ref{prop:expectation-opinion-fixed} and Lemma~\ref{lem:invertible-geometric}, now it is enough to show that $(qW)^K\to 0$ as $K\to\infty$, since then
    \[
    \lim_{k\to\infty}A_K
    = \lim_{K\to\infty}(I-qW)^{-1}\big(I-(qW)^K\big)
    = (I-qW)^{-1}.
    \]
    Let $\lambda$ be any eigenvalue of $qW$ with eigenvector $v\neq 0$. Then
    \[
    (qW)v=\lambda v
    \quad\Rightarrow\quad
    \|(qW)v\|_\infty=\|\lambda v\|_\infty=|\lambda|\,\|v\|_\infty.
    \]
    On the other hand, by the definition of the induced norm,
    \[
    \|(qW)v\|_\infty \le \|qW\|_\infty\,\|v\|_\infty.
    \]
    Dividing by $\|v\|_\infty>0$ gives $|\lambda|\le \|qW\|_\infty<1$, and therefore
    \[
    \rho(qW)=\max_i|\lambda_i(qW)|<1.
    \]
    This implies $(qW)^K\to 0$ as $K\to\infty$.
    Now, taking limits in Proposition~\ref{prop:expectation-opinion-fixed} yields
    \begin{align*}
        \lim_{K\to\infty}\mathbb{E}[X_K^\text{all}]
        &= (1-q)(I-qW)^{-1}x_0 + p(I-qW)^{-1}(u_0-x_0)\\
        &= (I-qW)^{-1}\Big((1-q)x_0+p(u_0-x_0)\Big).
    \end{align*}    
\end{proof}

\subsection{Proof of Corollary~\ref{cor:expectation-opinion-mixed-fixed-infinity}}\label{app:expectation-opinion-mixed-fixed-infinity}

\begin{proof}

Under Assumption~\ref{ass:no-outgoing-s}, after reindexing the agents if necessary, the matrix \(W\) can be written as
\[
W=
\begin{bmatrix}
W_{-s,-s} & 0\\
w_{s,-s} & w_{ss}
\end{bmatrix}.
\]
Since all agents \(i\neq s\) always click, their dynamics are deterministic and satisfy
\[
X_{-s,K}^{(1)}=\alpha x_{-s,0}+\beta W_{-s,-s}X_{-s,K-1}^{(1)}+(1-\zeta)u\mathbf 1_{-s}.
\]
Also, by Assumption~\ref{ass:no-outgoing-s}, the opinions of these \(n-1\) agents do not depend on the opinion of agent \(s\). Therefore, we can apply
Corollary~\ref{cor:expectation-opinion-fixed-infinity} to this subsystem of \(n-1\) agents with \(p=1-\zeta\) and \(q=\beta\). Hence,
\[
x_{-s}^\star = \lim_{K\to\infty}\mathbb{E}[X_{-s,K}^{(1)}]
=
(I-\beta W_{-s,-s})^{-1}
\Big(\alpha x_{-s,0}+(1-\zeta)u\mathbf 1_{-s}\Big).
\]
Now, we consider agent \(s\). Then,
\(
p=(1-\zeta)\gamma_0\) and  \(q=b+(\beta-b)\gamma_0.
\)
Taking expectation in the update of agent \(s\), and using the fact that the current click decision is independent of the past opinions, gives
\begin{align*}
\mathbb E[X_{s,K}^{(1)}]&=(1-p-q)x_{s,0}+p u\\
&\quad +q w_{s,-s}\mathbb E[X_{-s,K-1}^{(1)}]+q w_{ss}\mathbb E[X_{s,K-1}^{(1)}].
\end{align*}
Taking the limit as \(K\to\infty\), and using
\(
\lim_{K\to\infty}\mathbb E[X_{-s,K-1}^{(1)}]= x_{-s}^\star,
\)
we get
\begin{align*}
x_{s,\infty}^{(1)} &= \lim_{K\to\infty}\mathbb{E}[X_{s,K}^{(1)}]
=(1-p-q)x_{s,0}+p u\\
&\quad +q w_{s,-s}x_{-s}^\star+q w_{ss}x_{s,\infty}^{(1)}.
\end{align*}
Therefore,
\[
(1-qw_{ss})x_{s,\infty}^{(1)}=(1-p-q)x_{s,0}+p u+q w_{s,-s}x_{-s}^\star.
\]
Since \(q<1\) and \(w_{ss}\le 1\), we have \(1-qw_{ss}>0\). Hence,
\[
x_{s,\infty}^{(1)}=(1-qw_{ss})^{-1}
\Big((1-p-q)x_{s,0}+pu+q w_{s,-s}x_{-s}^\star
\Big).
\]
Therefore,
\[
\lim_{K\to\infty}\mathbb{E}\!\left[X_K^{(1)}\right]
=
\begin{bmatrix}
x_{-s}^\star\\[1mm]
x_{s,\infty}^{(1)}
\end{bmatrix}.
\]
\end{proof}

\subsection{Proof of Proposition~\ref{prop:gamma-convergence}}\label{app:gamma-convergence}

\begin{proof}

    \textbf{Approach 1.}
    First, we know under the assumption $\left|e_s^\top\Big((I-\beta W)^{-1}(\alpha x_0+(1-\zeta)u_0)-x_0\Big)\right| > \delta$ and initial probability clicking of 1 for all agents, agent $s$'s opinion is pushed at least $\delta$ away from its innate opinion ($|x_s^\star-x_{s,0}|>\delta$), and therefore triggers a reduction in probability of clicking. Now, we need to find the number of consecutive clicks ($M$) by agent $s$ that forces $|x_{s,t+M}-x_{s,0}|\ge \delta$ starting from time $t$. Throughout these $M$ time steps, the opinion of the population is given by
    \[
    x_{t+M}=(\beta W)^M x_t+\sum_{j=0}^{M-1}(\beta W)^j r,
    \]
    where all agents always click, $r:=\alpha x_0+(1-\alpha-\beta)u_0$, and $x^\star=(I-\beta W)^{-1}r$. Using $\sum_{j=0}^{M-1}(\beta W)^j r=(I-(\beta W)^M)x^\star$ yields
    \[
    x_{t+M}-x^\star=(\beta W)^M(x_t-x^\star).
    \]
    Considering the opinion of agent $s$ and applying the induced $\|\cdot\|_\infty$ norm (with $\|W\|_\infty=1$) gives
    \begin{align*}
        |x_{s,t+M}-x_s^\star|
        \le \|x_{t+M}-x^\star\|_\infty
        &\le \|\beta W\|_\infty^M \|x_t-x^\star\|_\infty\\
        &\le \beta^M \|x_t-x^\star\|_\infty .
    \end{align*}
    Since $x_t\in[-1,1]^n$ for all $t$ (Lemma~\ref{lem:convex-combination-opinion}), we have $\|x_t\|_\infty\le 1$, and hence using triangle inequality
    $\|x_t-x^\star\|_\infty\le \|x_t\|_\infty+\|x^\star\|_\infty \le 1+\|x^\star\|_\infty$ for all $t$. Using the reverse triangle inequality $|a+b|\ge |a|-|b|$ with $a:=x_s^\star-x_{s,0}$ and $b:=x_{s,t+M}-x_s^\star$, we get
    \begin{align*}
        |x_{s,t+M}-x_{s,0}|
        &=|(x_s^\star-x_{s,0})+(x_{s,t+M}-x_s^\star)|\\
        &\ge |x_s^\star-x_{s,0}|-|x_{s,t+M}-x_s^\star|\\
        &\ge |x_s^\star-x_{s,0}|-\beta^M(1+\|x^\star\|_\infty).
    \end{align*}

    To find $M$, we should have $|x_s^\star-x_{s,0}|-\beta^M(1+\|x^\star\|_\infty) \geq \delta$. Therefore, the finite integer $M$ is as follows:
    \[
    M:=\min\Big\{m\in\mathbb{N}:\ \beta^m(1+\|x^\star\|_\infty)\le |x_s^\star-x_{s,0}|-\delta\Big\}.
    \]
    
    Now let \(D(k):=\sum_{t=1}^{k}\mathbf 1\{|x_{s,t}-x_{s,0}|\ge \delta\}\), be the deviation counter and denote the number of times the threshold has been triggered up to time $k$, so that \(\gamma_{s,k}=\gamma_{s,0}\,\kappa^{-D(k)}=\kappa^{-D(k)}.\)

    To prove $\gamma_{s,k}\to 0$, it is enough to show that $D(k)\to\infty$ almost surely. To this end, we define the event that after a finite time $K$, the threshold is never triggered again and the number of reductions remains fixed at $d\in\mathbb{N}$, that is
    \[
    F_d:=\Big\{\exists K<\infty\ \text{s.t.}\ D(k)=d\ \text{for all }k\ge K\Big\}.
    \]
    Under the event $F_d$, for all $k\ge K$, we have $\gamma_{s,k}=p_d:= \;\kappa^{-d} > 0$.
    To show that $\mathbb{P}(F_d)=0$, let $\{U_k\}_{k\ge 0}$ be i.i.d. random variables uniformly distributed on $[0,1]$, and we use them to generate the clicks,
    \[
    c_{s,k} := \mathbf{1}\{U_k \le \gamma_{s,k}\},
    \]
    and also partition the time steps after $K$ into disjoint blocks of length $M$:
    \[
    B_m:=\{K+mM,\ K+mM+1,\ \dots,\ K+mM+M-1\}\]
    for $m=0,1,2,\dots$.
    For each block, we can also define the event under which agent $s$ clicks throughout block $m$, which is:
    \[
    E_m:=\bigcap_{k\in B_m}\{U_k\le p_d\}.
    \]
    Since the blocks $\{B_m\}$ are disjoint and the uniforms $\{U_k\}$ are i.i.d., the events $\{E_m\}_{m\ge0}$ are independent and satisfy
    \[
    \mathbb{P}(E_m)=p_d^M>0.
    \]
    Hence $\sum_{m=0}^{\infty}\mathbb{P}(E_m)=\infty$. By the (second) Borel--Cantelli lemma, $\mathbb{P}(E_m\ \text{i.o.})=1$.
    Now, suppose that $F_d$ occurs. Then, there exists a finite time $K$ such that $\gamma_{s,k}=p_d$ for all $k\ge K$. Since the events $E_m$ occur infinitely often almost surely, at least one such block occurs after this time $K$. On that block, agent $s$ clicks for $M$ consecutive steps, which forces $|x_{s,\cdot}-x_{s,0}|\ge\delta$ at the end of the block, and therefore triggers another reduction. This increases $D(k)$ beyond $d$, contradicting the definition of $F_d$. Therefore, $\mathbb{P}(F_d)=0$ for every $d\in\mathbb{N}$.
    Since $D(k)$ is integer and nondecreasing, if it does not diverge then it must eventually be constant.
    Thus $\{D(k)\ \text{does not diverge}\}\subseteq \bigcup_{d\ge0}F_d$, and hence
    \[
    \mathbb{P}\big(D(k)\ \text{does not diverge}\big)\le \sum_{d\ge0}\mathbb{P}(F_d)=0.
    \]

    This implies $D(k)\to\infty$ almost surely, and hence $\gamma_{s,k}=\kappa^{-D(k)}\to 0$ almost surely.

    \textbf{Approach 2.} The second approach uses the intuition from \emph{positive crossing probability}. We prove the result by showing that the clicking probability cannot stay bounded away from zero.
    We define the event
    \[
    H:=\left\{\omega:\exists \,\varepsilon>0 \text{ such that } \gamma_{s,k}(\omega)\ge \varepsilon \text{ for infinitely many } k\right\}
    \]

    Our goal is to show $\mathbb{P}(H)=0$ (the probability of the trajectories in which the clicking probability stays bounded away from zero is zero), which implies $\gamma_{s,k}\to 0$ almost surely.
    For each \(\ell\in\mathbb N\), let
    \[
    H_\ell := \left\{ \omega : \gamma_{s,k}(\omega) \geq \tfrac{1}{\ell} \text{ for infinitely many } k \right\}.
    \]
    Since $H \subseteq \bigcup_{\ell=1}^\infty H_\ell$, it is enough to show that \( \mathbb{P}(H_\ell) = 0 \) for every \(\ell\). Let $R:=\{x:\ |x-x_{s,0}|<\delta, x \in [0,1]\}$ be the tolerance region where no reduction is triggered when the opinion lies in it. For a fixed $\ell\in\mathbb{N}$, under the event $H_\ell$, the process must remain in $R$ infinitely often while $\gamma_{s,k}\ge \frac1\ell$. Otherwise, if
    $|X_{s,k}-x_{s,0}|\ge\delta$ occurs infinitely often with
    $\gamma_{s,k}\ge \frac1\ell$, then infinitely many reductions would be
    triggered, which would eventually force \(\gamma_{s,k}<1/\ell\), contradicting \(H_\ell\).
    
    Considering the same definition of $M$ from Approach~1, under the assumption $|x_s^\star-x_{s,0}|>\delta$, we define the stopping times recursively by $\tau_0:=0$ and for $j\ge1$, \[\tau_j:=\inf\left\{k\ge \tau_{j-1}+M:\ |X_{s,k}-x_{s,0}|<\delta,\ \gamma_{s,k}\ge \frac1\ell \right\},\]

    where $\tau_1< \tau_2< \cdots$, and $\tau_j$ is the j-th time the process is in $R$ with $\gamma_{s,k}$ at least $\frac{1}{\ell}$ after the previous $j-1$ blocks of $M$. Under the event $H_\ell$, there are infinitely many such time indices, i.e., $\tau_j<\infty$ for all $j$.

    Now we use the realization of Bernoulli clicks, where $U_k\stackrel{\text{i.i.d.}}{\sim}\mathrm{Unif}[0,1]$ and the clicks are generated as \(c_{s,k}=\mathbf{1}\{U_k\le \gamma_{s,k}\}.\)
    
    Conditioned on $\mathcal{F}_{\tau_j}$ (the history up to time $\tau_j$), if \(\tau_j<\infty\), we have
    $\gamma_{s,\tau_j}\ge \tfrac{1}{\ell}$. As long as no reduction has occurred during the next $M$ steps, the clicking probabilities stay at least $\tfrac{1}{\ell}$. Therefore, the probability that agent $s$ clicks for the next $M$ steps is at least
    \(q_\ell := \Big(\tfrac1\ell\Big)^M>0\)
    (exiting in fewer than \(M\) clicks only increases this probability).

    By the definition of $M$, this event forces the process to exit $R$ within that window and hence triggers a reduction.

    Now, let \(I_j\) be the event that no exit from \(R\) occurs during the \(j\)-th window. Then, on \(\{\tau_j<\infty\}\),
    \[\mathbb P(I_j\mid \mathcal F_{\tau_j})\le 1-q_\ell.\]
    
    Now, we consider the event that the first \(J\) trial windows all succeed, that is, \(\bigcap_{j=1}^J I_j\). Let
    \(A_{j-1}:=\bigcap_{i=1}^{j-1}I_i.
    \)
    Since the first $j-1$ windows are completed before the $j$-th window starts, we have $A_{j-1}\in\mathcal F_{\tau_j}$. Therefore, using the tower property of conditional expectation, for any \(j\) with \(\mathbb P(A_{j-1})>0\),
    \begin{align*}
        \mathbb P(I_j\mid A_{j-1})
        &= \frac{\mathbb E[\mathbf 1_{A_{j-1}}\mathbf 1_{I_j}]}{\mathbb P(A_{j-1})}=\frac{\mathbb E\!\left[\mathbf 1_{A_{j-1}}\mathbb E[\mathbf 1_{I_j}\mid\mathcal F_{\tau_j}]\right]}
        {\mathbb P(A_{j-1})}\\
        &=\frac{\mathbb E\!\left[\mathbf 1_{A_{j-1}}\mathbb P(I_j\mid\mathcal F_{\tau_j})\right]}
        {\mathbb P(A_{j-1})}\le\frac{\mathbb E\!\left[\mathbf 1_{A_{j-1}}(1-q_\ell)\right]}{\mathbb P(A_{j-1})}\\
        &=1-q_\ell.
    \end{align*}
    If $\mathbb P(A_{j-1})=0$, the bound is trivial. Hence, by the chain rule and the bound above,
    \[
    \mathbb{P}\Big(\bigcap_{j=1}^J I_j\Big)=
    \prod_{j=1}^J\mathbb P(I_j\mid A_{j-1})\le
    \prod_{j=1}^J(1-q_\ell)
    =(1-q_\ell)^J.
    \]
    Letting $J\to\infty$, we get
    \[\mathbb{P}\Big(\bigcap_{j=1}^\infty I_j\Big)
    \le \lim_{J\to\infty}(1-q_\ell)^J = 0.
    \]
    Hence, with probability one, at least one of these windows leads to an exit from $R$, and therefore to a reduction of $\gamma_{s,k}$.
    The same argument can be repeated after each reduction. As long as $H_\ell$
    holds, there are still infinitely many times at which the process is in $R$
    and $\gamma_{s,k}\ge \frac1\ell$. Hence, reductions can happen infinitely
    often on $H_\ell$. This is a contradiction, because after sufficiently many reductions, \(\gamma_{s,k}<\frac1\ell,\)
    which contradicts the fact that \(\gamma_{s,k}\ge 1/\ell\) infinitely often. Therefore, $\mathbb{P}(H_\ell)=0$.

    Since this holds for every $\ell\in\mathbb{N}$, we conclude
    \[
    \mathbb{P}(H) \leq \mathbb{P}\left(\bigcup_{\ell=1}^\infty H_\ell\right) \leq \sum_{\ell=1}^{\infty} \mathbb{P} (H_\ell) = 0,
    \]
    so $\limsup_{k\to\infty}\gamma_{s,k}=0$ almost surely. Because $\gamma_{s,k}\ge0$, this implies
    $\gamma_{s,k}\to0$ almost surely.
\end{proof}

\subsection{Proof of Corollary~\ref{cor:expectation-opinion-adaptive-infinity}}\label{app:expectation-opinion-adaptive-infinity}

\begin{proof}

The proof follows the same idea as Corollary~\ref{cor:expectation-opinion-mixed-fixed-infinity}. Under Assumption~\ref{ass:no-outgoing-s}, the opinions of the other agents are not influenced by agent \(s\). Moreover, these agents follow the same policy
as in Corollary~\ref{cor:expectation-opinion-mixed-fixed-infinity}, that is, they
always click. Therefore,
\[
\lim_{K\to\infty}\mathbb E[X_{-s,K}^{(2)}]=x_{-s}^\star=(I-\beta W_{-s,-s})^{-1}
\Big(\alpha x_{-s,0}+(1-\zeta)u\mathbf 1_{-s}\Big).
\]

It remains to characterize the limiting expected opinion of agent \(s\). From the compact form of the dynamics \eqref{eq:user-model-PQ}, the update of agent \(s\) can be written as
\begin{align*}
    X_{s,K+1}^{(2)}&=(1-Q_{s,K}-P_{s,K})x_{s,0}\\
    &\quad+Q_{s,K}\Big(w_{s,-s}X_{-s,K}^{(2)}+w_{ss}X_{s,K}^{(2)}\Big) +P_{s,K}u,
\end{align*}
where \(Q_{s,K}=b+(\beta-b)c_{s,K}\) and \(P_{s,K}=(1-\zeta)c_{s,K}\). Let \(\mathcal F_K\) denote the history of the process before the click decision at time \(K\), including the opinions \(\{x_t\}_{t=0}^{K}\), the past click decisions \(\{c_t\}_{t=0}^{K-1}\), and the fixed recommendation \(u_0\). Given this history, the value of \(\gamma_{s,K}\) is determined by the adaptive policy, and agent \(s\) clicks with probability \(\gamma_{s,K}\). Therefore,
\[
\mathbb E[c_{s,K}\mid \mathcal F_K]=\gamma_{s,K}.
\]
By the law of total expectation,
\[
\mathbb E[c_{s,K}]=\mathbb E\!\left[\mathbb E[c_{s,K}\mid \mathcal F_K]\right]=\mathbb E[\gamma_{s,K}].
\]
Since Proposition~\ref{prop:gamma-convergence} gives \(\gamma_{s,K}\to0\) almost surely and \(0\le \gamma_{s,K}\le1\), it follows that \(\mathbb E[c_{s,K}]=\mathbb E[\gamma_{s,K}]\to 0.\)
Therefore, the click-dependent terms in the update of agent \(s\) vanish in expectation as \(K\to\infty\). Hence, in the limit, the expected opinion of agent \(s\) reduces to the no click update:
\[
x_{s,\infty}^{(2)}
=
(1-b)x_{s,0}
+
b\Big(w_{s,-s}x_{-s}^\star+w_{ss}x_{s,\infty}^{(2)}\Big).
\]
Equivalently,
\[
(1-bw_{ss})x_{s,\infty}^{(2)}=(1-b)x_{s,0}+b w_{s,-s}x_{-s}^\star.
\]
Since \(b<1\) and \(w_{ss}\le 1\), we have \(1-bw_{ss}>0\). Thus,
\[
x_{s,\infty}^{(2)}=(1-bw_{ss})^{-1}
\Big((1-b)x_{s,0}+b\,w_{s,-s}x_{-s}^\star\Big).
\]
Combining this with the limit of the passive agents gives
\[
\lim_{K\to\infty}\mathbb E[X_K^{(2)}]=\begin{bmatrix}
x_{-s}^\star\\[1mm]
x_{s,\infty}^{(2)}
\end{bmatrix},
\]
which completes the proof.
\end{proof}

\subsection{Proof of Lemma~\ref{lem:compare-utility-infinite}}\label{app:compare-utility-infinite}

We begin the proof of this lemma by introducing a new lemma, where we use Assumption~\ref{ass:sign-preservation}, to simplify the deviation $|X_{s,k}-x_{s,0}|$ in the utility for a reactive agent $s$.

\begin{lemma}\label{lem:sigma}
Under Assumption~\ref{ass:sign-preservation}, at each time step $k\geq 0$, the opinion of agent $s$ stays on the same side of its innate opinion, i.e.,
\(
x_{s,0}\le X_{s,k},
\)
when \(
x_{s,0}\le u,
\)
and
\(
x_{s,0}\ge X_{s,k},
\)
when
\(
x_{s,0}\geq u,
\)
for all $k \geq 0$. Consequently,
\[
|X_{s,k}-x_{s,0}|=\sigma_s\bigl(X_{s,k}-x_{s,0}\bigr),\qquad \forall k\ge 0,
\]
where $\sigma_s=\operatorname{sign}(u-x_{s,0})$. This property is assumed to hold for every realization of the click decisions. In particular, it applies under both policy~\ref{alg:policy-one} and policy~\ref{alg:policy-two}.
\end{lemma}

Based on the opinion dynamics~\eqref{eq:user-model-componentwise}, $X_{s,k}$ is always a convex combination of the innate opinion $x_{s,0}$, the neighbors' influence $(WX_k)_s$,
and (when the agent clicks) the recommendation $u$.
Under Assumption~\ref{ass:sign-preservation}, these quantities remain on the same side of $x_{s,0}$ for all $k$, which implies that $X_{s,k}$ cannot cross its innate opinion $x_{s,0}$.

Now, we turn to the proof of Lemma~\ref{lem:compare-utility-infinite}.

\begin{proof}
    Using the utility definition in \eqref{eq:agent-utility}, the assumption \(R^A(|x_{i,k}-u_{i,k}|)=1\), and Lemma~\ref{lem:sigma}, for each policy \(\texttt{p}\in\{1,2\}\), we have
    \[\mathbb E\!\left[U_s^{(\texttt{p})}(h_{s,K}^{(A)})\right]=\lambda \frac{1}{K}\sum_{k=0}^{K-1}\mathbb E[c_{s,k}^{(\texttt{p})}]-(1-\lambda)\sigma_s \big(\mathbb E[X_{s,K}^{(\texttt{p})}]-x_{s,0}\big).\]
    
    The fixed clicking policy~\ref{alg:policy-one} gives
    \(\mathbb E[c_{s,k}^{(1)}]=\gamma_0\) for all \(k\). Therefore,
    \(\frac{1}{K}\sum_{k=0}^{K-1}\mathbb E[c_{s,k}^{(1)}]=\gamma_0.\)
    Using Corollary~\ref{cor:expectation-opinion-mixed-fixed-infinity} and taking
    \(K\to\infty\), we obtain
    \[
    \lim_{K\to\infty}\mathbb{E}\!\left[U_s^{(1)}(h_{s,K}^{(A)})\right]=\lambda\,\gamma_0
    -(1-\lambda)\,\sigma_s\Big(x_{s,\infty}^{(1)}-x_{s,0}\Big).\]

    For the adaptive decreasing clicking policy~\ref{alg:policy-two}, the proof of Corollary~\ref{cor:expectation-opinion-adaptive-infinity}, implies $\mathbb E[c_{s,k}^{(2)}]\to 0$. Thus,
    \(\lim_{K\to\infty}\frac{1}{K}\sum_{k=0}^{K-1}\mathbb E[c_{s,k}^{(2)}]=0.\)

    Moreover, by Corollary~\ref{cor:expectation-opinion-adaptive-infinity},
    \(\lim_{K\to\infty}\mathbb E[X_{s,K}^{(2)}]=x_{s,\infty}^{(2)}.\) Taking the limit of the expected utility, we obtain
    \[\lim_{K\to\infty}\mathbb E\!\left[U_s^{(2)}(h_{s,K}^{(A)})\right]=-(1-\lambda)\sigma_s \Big(x_{s,\infty}^{(2)}-x_{s,0}\Big).\]
\end{proof}


\subsection{Proof of Proposition~\ref{prop:finite-adaptive-better-than-fixed}}\label{app:finite-adaptive-better-than-fixed}

\begin{proof}
    Under Assumption~\ref{ass:no-outgoing-s}, opinion of passive agents \(\{X_{-s,k}\}_{k\ge 0}\) is deterministic and follows the all click recursion on
    \((W_{-s,-s},x_{-s,0})\) (equivalently, Proposition~\ref{prop:expectation-opinion-fixed} with \(\gamma_0=1\)).
    
    Let $\{X^{(1)}_{s,k}\}$ and $\{X^{(2)}_{s,k}\}$ denote the opinion processes of agent $s$ under the fixed policy~\ref{alg:policy-one} and the adaptive policy~\ref{alg:policy-two}, respectively. Since $R^A(\cdot) = 1$,
    \begin{align*}\label{eq:gap-start}
        &\mathbb{E}[U^{(2)}]-\mathbb{E}[U^{(1)}] =-\lambda\Big(\gamma_0-\bar\gamma_K\Big)\\
        &\quad +(1-\lambda)\Big(\mathbb{E}[|X^{(1)}_{s,K}-x_{s,0}|]-\mathbb{E}[|X^{(2)}_{s,K}-x_{s,0}|]\Big),
    \end{align*}
    where $\bar\gamma_K:=\frac{1}{K}\sum_{k=0}^{K-1}\mathbb{E}[\gamma_{s,k}]$ is the adaptive average clicking probability.

    Using Lemma~\ref{lem:sigma}, for \(\texttt{p}\in\{1,2\}\), we have \(
    \mathbb{E}[|X^{(\texttt{p})}_{s,K}-x_{s,0}|]=\sigma_s(\mathbb{E}[X^{(\texttt{p})}_{s,K}]-x_{s,0}),\)
    where $\sigma_s=\operatorname{sign}(u-x_{s,0})$. Therefore, the mean drift gap is 
    \begin{align*}
        d_K &:= \mathbb{E}[|X^{(1)}_{s,K}-x_{s,0}|]-\mathbb{E}[|X^{(2)}_{s,K}-x_{s,0}|]\\
        &=\sigma_s\Big(\mathbb{E}[X^{(1)}_{s,k}]-\mathbb{E}[X^{(2)}_{s,k}]\Big).
    \end{align*}
    Then, the gap in expected utilities is
    \[
    \mathbb{E}[U^{(2)}]-\mathbb{E}[U^{(1)}]=-\lambda(\gamma_0-\bar\gamma_K)+(1-\lambda)d_K.
    \]
    To find the condition under which adaptive decreasing policy has higher expected utility, it suffices to ensure 
    \(\lambda(\gamma_0-\bar\gamma_K) < (1-\lambda)d_K\).
    
    We first upper bound \(\gamma_0-\bar\gamma_K\) using \(M\). Under Lemma~\ref{lem:sigma}, the deviation \(\sigma_s(X_{s,k}-x_{s,0})\) increases by a click update and decreases by a no click update. Therefore, among all sample paths up to time \(k\), the largest possible deviation \(|X_{s,k}-x_{s,0}|\) is attained by the all click path of length \(k\). By the geometric convergence bound for the all click recursion (as used in the proof of Proposition~\ref{prop:gamma-convergence})
    and the definition of \(M\), the deviation is still \(<\delta\) for all \(k<M\). Hence no reduction can be triggered before time \(M\),
    so \(\gamma_{s,k}=\gamma_0\) for \(k=0,\dots,M-1\). After time \(M\), each time a reduction happens, the click probability is divided by \(\kappa\), so for \(j=1,\dots,K-M\),
    the smallest possible value at time \(M-1+j\) is achieved when a reduction happens at every step after \(M\), hence \[\gamma_{s,M-1+j}\ \ge\ \frac{\gamma_0}{\kappa^{\,j}}.\]
    Taking expectations and averaging gives the lower bound
    \[
    \bar\gamma_K \ge \frac{\gamma_0}{K}\Big(M+\sum_{j=1}^{K-M}\kappa^{-j}\Big)
    = \frac{\gamma_0}{K}\Big(M+\frac{1-\kappa^{-(K-M)}}{\kappa-1}\Big),
    \]
    and therefore an upper bound on \(\gamma_0-\bar\gamma_K\).

    To lower bound \(d_K\), let the fixed policy mean be $m_k:=\mathbb{E}[X^{(1)}_{s,k}]$ and the adaptive policy mean be \(\widetilde m_k:=\mathbb{E}[X^{(2)}_{s,k}]\). Since \(X_{-s,k}\) is deterministic and \(p(\gamma)=(1-\zeta)\gamma\), \(q(\gamma)=b+(\beta-b)\gamma\),
    taking expectations in the one step update~\eqref{eq:user-model-PQ} for agent $s$ under policy~\ref{alg:policy-one} and~\ref{alg:policy-two} yields $m_{k+1}$ and $\widetilde m_{k+1}$, respectively, that is
    \begin{align*}
        m_{k+1} &= q(\gamma_0)\big(w_{ss}m_k+w_{s,-s}X_{-s,k}\big)\\
        \quad & +\big(1-q(\gamma_0)-p(\gamma_0)\big)x_{s,0}+p(\gamma_0)u,\\
        \widetilde m_{k+1} &=w_{ss}\,\mathbb{E}\!\big[q(\gamma_{s,k})\,X^{(2)}_{s,k}\big]
        +w_{s,-s}X_{-s,k}\,\mathbb{E}\!\big[q(\gamma_{s,k})\big]\\
        &\quad
        +x_{s,0}\Big(1-\mathbb{E}\!\big[q(\gamma_{s,k})\big]-\mathbb{E}\!\big[p(\gamma_{s,k})\big]\Big)+u\,\mathbb{E}\!\big[p(\gamma_{s,k})\big],
    \end{align*}
    where $m_0=\widetilde m_0=x_{s,0}$.
    
    Since \(p(\cdot)\) and \(q(\cdot)\) are affine, linearity of expectation implies
    \(\mathbb{E}[p(\gamma_{s,k})]=p(\mathbb{E}[\gamma_{s,k}])\) and \(\mathbb{E}[q(\gamma_{s,k})]=q(\mathbb{E}[\gamma_{s,k}])\). Now, we define the random click probability gap \(\widetilde\Delta_k:=\gamma_0-\gamma_{s,k}\ge 0\) and its mean
    \(\Delta_k:=\mathbb{E}[\widetilde\Delta_k]=\gamma_0-\mathbb{E}[\gamma_{s,k}]\ge 0\).
    Subtracting the two mean recursions, multiplying by \(\sigma_s\), and using the affine forms of \(p,q\) yields
    \[
    d_{k+1}=\rho\,d_k + \sigma_s B_k\,\Delta_k + \sigma_s\varepsilon_k,
    \]
    where
    \begin{align*}
        \rho&:=(b+(\beta-b)\gamma_0)w_{ss}\\
        B_k &:= (\beta-b)\big(w_{s,-s}X_{-s,k}-x_{s,0}\big) + (1-\zeta)(u-x_{s,0}),\\
        \varepsilon_k &:= w_{ss}(\beta-b)\,\mathbb{E}\big[\widetilde\Delta_k\,X^{(2)}_{s,k}\big].
    \end{align*}

    Using $|X^{(2)}_{s,k}|\le 1$ and $\widetilde\Delta_k\ge 0$, we have $
    \mathbb{E}\big[\widetilde\Delta_k X^{(2)}_{s,k}\big]\ge -\mathbb{E}[\widetilde\Delta_k]=-\Delta_k.$
    Moreover, since $b=\frac{\beta}{\zeta}$, $\alpha \geq \beta$ and $0<\zeta\le 1$, we have $b\ge \beta$, hence $\beta-b\le 0$ and
    $|\beta-b|=b-\beta$. Therefore,
    \[\sigma_s\varepsilon_k = \sigma_s w_{ss}(\beta-b)\,\mathbb{E}\!\big[\widetilde\Delta_k\, X^{(2)}_{s,k}\big]\ge\ 
    -\,\sigma_s w_{ss}\,|\beta-b|\,\Delta_k.\]
    Substituting this lower bound in the definition of $d_k$, results in
    \[d_{k+1}\ \ge\ \rho\,d_k + \sigma_s\big(B_k-w_{ss}|\beta-b|\big)\Delta_k = \rho\,d_k + g_k\,\Delta_k,
    \]
    where
    \[g_k = \sigma_s(1-\zeta)\Big((u-x_{s,0})-b\big(w_{s,-s}X_{-s,k}-x_{s,0}+w_{ss}\big)\Big).\]
    This matches the definition in Proposition~\ref{prop:finite-adaptive-better-than-fixed}.
    
    Since \(d_0=0\), iterating yields
    \[
    d_K\ \ge\ \sum_{k=0}^{K-1}\rho^{K-1-k}g_k\,\Delta_k.
    \]
    By the explanation above, the first reduction cannot occur before time \(M\), and the proposition assumes at least one reduction occurs by time \(K\). 
    Hence for all \(k\ge M\), the click probability has been reduced at least once, so \(\gamma_{s,k}\le \frac{\gamma_0}{\kappa}\), and thus
    \[
    \Delta_k=\gamma_0-\mathbb{E}[\gamma_{s,k}] \ \ge\ \gamma_0\Big(1-\kappa^{-1}\Big),
    \qquad k=M,\dots,K-1.
    \]
    Dropping the nonnegative terms for \(k<M\) (which is the same under both policies) gives
    \[
    d_K\ \ge\ \gamma_0(1-\kappa^{-1})\sum_{k=M}^{K-1}\rho^{K-1-k}g_k
    =\gamma_0(1-\kappa^{-1})G_M.
    \]

    Combining the sufficient condition  \(\lambda(\gamma_0-\bar\gamma_K) < (1-\lambda)d_K\), with the clicking upper bound and the drift lower bound yields
    \[\lambda \gamma_0 \Big(1 - \frac{1}{K}\Big(M+\frac{1-\kappa^{-(K-M)}}{\kappa-1}\Big)\Big) <
    (1-\lambda)\gamma_0(1-\kappa^{-1})G_M.\]
    Canceling \(\gamma_0 > 0\),
    we obtain
    \[
    \lambda \;<\; \frac{1}{\,1+\dfrac{K-M-\frac{1-\kappa^{-(K-M)}}{\kappa-1}}{K(1-\kappa^{-1})\,G_M}\,},
    \]
    This is exactly the condition in Proposition~\ref{prop:finite-adaptive-better-than-fixed}, concluding the proof.
\end{proof}

\end{document}